\documentclass[aps,prd,10pt,preprintnumbers,superscriptaddress,amsmath,twocolumn,showpacs,nofootinbib]{revtex4-2}

\usepackage[dvipsnames]{xcolor}
\usepackage[sort&compress]{natbib}
\usepackage{color,hhline,psfrag,rotating,amssymb}
\usepackage[colorlinks=true,linkcolor=PineGreen,filecolor=PineGreen,urlcolor=PineGreen,citecolor=PineGreen,plainpages=false]{hyperref}
\usepackage[hang,nooneline]{subfigure}
\usepackage{dcolumn}

\usepackage{graphicx}
\usepackage{revsymb}
\usepackage{dcolumn}

\usepackage{times}

\begin{document}
\preprint{FERMILAB-PUB-22-450-T}
\title{Windows on the hadronic vacuum polarisation contribution to the muon anomalous magnetic moment}
\author{C.~T.~H.~Davies}\email{christine.davies@glasgow.ac.uk}
\affiliation{SUPA, School of Physics and Astronomy, University of Glasgow, Glasgow, G12 8QQ, UK}
\author{C.~DeTar} 
\affiliation{Department of Physics and Astronomy, University of Utah, Salt Lake City, Utah 84112, USA}
\author{A.~X.~El-Khadra} 
\affiliation{Department of Physics, University of Illinois, Urbana, Illinois 61801, USA}
\affiliation{Illinois Centre for Advanced Studies of the Universe, University of Illinois, Urbana, Illinois 61801, USA}
\author{Steven~Gottlieb} 
\affiliation{Department of Physics, Indiana University, Bloomington, Indiana 47405, USA}
\author{D.~Hatton} 
\affiliation{SUPA, School of Physics and Astronomy, University of Glasgow, Glasgow, G12 8QQ, UK}
\author{A.~S.~Kronfeld} 
\affiliation{Fermi National Accelerator Laboratory, Batavia, Illinois 60510, USA}
\author{S.~Lahert} 
\affiliation{Department of Physics, University of Illinois, Urbana, Illinois 61801, USA}
\author{G.~P.~Lepage}\email{g.p.lepage@cornell.edu}
\affiliation{Laboratory for Elementary-Particle Physics, Cornell University, Ithaca, New York 14853, USA}
\author{C.~McNeile}
\affiliation{Centre for Mathematical Sciences, University of Plymouth, PL4 8AA, UK}
\author{E.~T.~Neil}
\affiliation{Department of Physics, University of Colorado, Boulder, Colorado 80309, USA}
\author{C.~T.~Peterson}
\affiliation{Department of Physics, University of Colorado, Boulder, Colorado 80309, USA}
\author{G.~S.~Ray}
\affiliation{Centre for Mathematical Sciences, University of Plymouth, PL4 8AA, UK}
\author{R.~S.~Van de Water} 
\affiliation{Fermi National Accelerator Laboratory, Batavia, Illinois 60510, USA}
\author{A.~Vaquero} 
\affiliation{Department of Physics and Astronomy, University of Utah, Salt Lake City, Utah 84112, USA}
             \collaboration{Fermilab Lattice, HPQCD, and MILC Collaborations}
        \homepage{http://www.physics.gla.ac.uk/HPQCD}
        \noaffiliation
\date{\today}

\begin{abstract}
An accurate determination  of the leading-order hadronic vacuum polarisation (HVP) contribution to the anomalous magnetic moment of the muon is critical to understanding the size and significance of any discrepancy between the Standard Model prediction and experimental results being obtained by the Muon g-2 experiment at Fermilab. The Standard Model prediction is currently based on a data-driven approach to the HVP using experimental results for $\sigma(e^+e^-\rightarrow\,\mathrm{hadrons})$. Lattice QCD aims to provide a result with similar uncertainty from calculated vector-vector correlation functions, but the growth of statistical and systematic errors in the $u/d$ quark correlation functions at large Euclidean time has made this difficult to achieve. We show that restricting the lattice contributions to a one-sided window $0<t<t_1$ can greatly improve lattice results while still capturing a large fraction of the total HVP. 
We illustrate this by comparing windowed lattice results based on the 2019 Fermilab Lattice/HPQCD/MILC HVP analysis with corresponding results obtained from the KNT19 analysis of $R_{e^+e^-}$ data.
For $t_1=1.5$\,fm, 70\% of the total HVP is contained within the window and our lattice result has an error of~0.7\%, only about twice as big as the error from the $e^+e^-$~analysis. We see a tension of 2.7$\sigma$ between the two results. With increased statistics in the lattice data the one-sided windows will allow stringent tests of lattice and $R_{e^+e^-}$ results that include a large fraction of the total HVP contribution.
\end{abstract}

\maketitle
\section{Introduction}

The anomalous magnetic moment of the muon, $a_{\mu}$, captures the impact on the properties of the muon of its interaction with the sea of virtual particles present in the deep subatomic world. $a_{\mu}$ is currently being measured to an unprecedented level of precision at the Muon g-2 experiment at Fermilab~\cite{Muong-2:2021ojo}. Comparison with the expectation from the Standard Model (SM)~\cite{Aoyama:2020ynm}, if it can be done well enough, has the potential to uncover the existence of new particles beyond those of the SM in the virtual sea. The existence of such particles would be signalled by a significant discrepancy in the value of $a_{\mu}$ between the SM expectation, including the effect of all known particles, and the experimental result. 

The first result from the Muon g-2 experiment~\cite{Muong-2:2021ojo} gives a new experimental average value for $a_{\mu}$ that is larger than the SM expectation~\cite{Aoyama:2020ynm}  by 25.1(5.9) $\times 10^{-10}$, showing a tantalising 4.2$\sigma$ tension. 

\begin{figure}
    \includegraphics[scale=0.9]{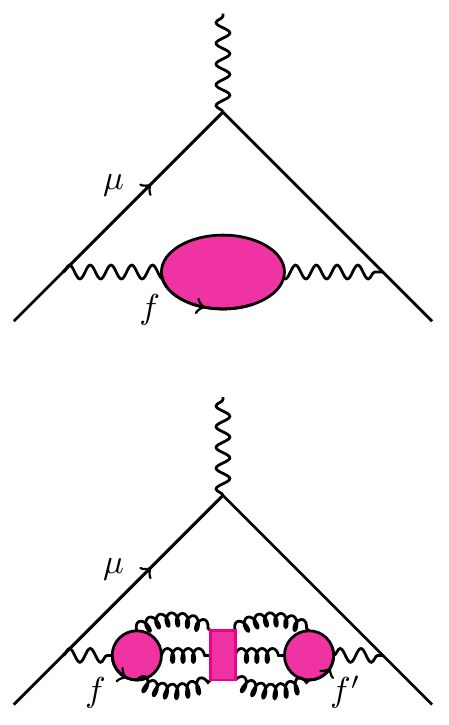}
    \caption{\label{fig:bubble} HVP contribution to $a_{\mu}$. The upper plot shows the quark-line connected contribution and the lower plot the quark-line disconnected contribution. Wavy lines are photons and curly lines are gluons. These contributions are taken to include all possible QED interactions within the strong interaction bubble. 
    }
\end{figure}

A key contribution to the SM value (which is based on Refs.~\cite{Aoyama:2012wk,Aoyama:2019ryr,Czarnecki:2002nt,Gnendiger:2013pva,Davier:2017zfy,Keshavarzi:2018mgv,Colangelo:2018mtw,Hoferichter:2019mqg,Davier:2019can,Keshavarzi:2019abf,Kurz:2014wya,Melnikov:2003xd,Masjuan:2017tvw,Colangelo:2017fiz,Hoferichter:2018kwz,Gerardin:2019vio,Bijnens:2019ghy,Colangelo:2019uex,Blum:2019ugy,Colangelo:2014qya}) is that from the leading-order hadronic vacuum polarisation contribution. We will denote this by the acronym HVP in what follows. This contribution is sizeable, second only to the dominant QED contribution which has a small uncertainty because it has been calculated through fifth order in the QED coupling, $\alpha$~\cite{Aoyama:2012wk,Aoyama:2019ryr}. The HVP is much harder to pin down because it involves strong interaction physics at low momentum scales. As shown in Fig.~\ref{fig:bubble}, it arises from a virtual quark bubble (or bubbles connected by gluons) inserted in a photon propagator. Calculation of the HVP contribution can be expressed as the integral over space-like $q^2$ of the vacuum polarisation function, $\hat{\Pi}(q^2)$, with a kernel function that emphasises small $|q^2|$ values of $\mathcal{O}(m_{\mu}^2)$~\cite{Lautrup:1971jf, deRafael:1993za, Blum:2002ii}. In the SM, the integral over $\hat{\Pi}(q^2)$ can be straightforwardly calculated in lattice QCD by working in coordinate space~\cite{Bernecker:2011gh, Chakraborty:2014mwa}, for an effective `first principles' approach. The primary quantities needed are the correlation functions between two electromagnetic current operators as a function of their time separation (summed over spatial coordinates at either end). Achieving small statistical and systematic uncertainties is challenging~\cite{Budapest-Marseille-Wuppertal:2017okr, RBC:2018dos, Giusti:2018mdh, Shintani:2019wai, FermilabLattice:2019ugu, Gerardin:2019rua}, however, for the dominant contributions where the current couples to $u$ or $d$ quarks. This will be discussed further below. 

At present the SM value for the HVP is taken from `data-driven' approaches that use the wealth of detailed experimental data for the cross-section for $e^+e^-\rightarrow\,$hadrons. The ratio of cross-sections for $e^+e^-\rightarrow\,$hadrons to that for $e^+e^-\rightarrow\mu^+\mu^-$, $R_{e^+e^-}$, is obtained as a function of centre-of-mass energy, $\sqrt{s}$, and related to $\hat{\Pi}$ for time-like $q^2$. The analytic structure of $\hat{\Pi}$ in the complex $q^2$-plane then allows the HVP to be determined from an integral over $s$ of $R_{e^+e^-}$ with a kernel function that emphasises small values of $\sqrt{s}$. 

The SM prediction for $a_{\mu}$ in Ref.~\cite{Aoyama:2020ynm} uses a data-driven evaluation of the HVP based on Refs.~\cite{Davier:2017zfy, Keshavarzi:2018mgv, Colangelo:2018mtw, Hoferichter:2019mqg,  Davier:2019can, Keshavarzi:2019abf} with a 0.6\% uncertainty. It is this result for the HVP ($693.1(4.0)\times 10^{-10}$) that yields the tension of 4.2$\sigma$ between the SM and experiment mentioned above. 

The recent lattice-QCD result for the HVP from the BMW collaboration~\cite{Borsanyi:2020mff} is the most complete to date and has an uncertainty of 0.8\%. It is, however, 2.1$\sigma$ higher than the combined data-driven HVP and yields a value for $a_{\mu}$ within 1.6$\sigma$ of the experimental value. The current level of uncertainty does not yet allow any clear conclusion on whether the two HVP values differ or not. Such a conclusion needs other independent lattice-QCD results with uncertainties that are improved to the level of the BMW result or better. 

More compelling in terms of a comparison between the lattice and data-driven approaches to the HVP is the calculation of a part of the HVP obtained by imposing a time-window  on the lattice correlation functions to remove the problematic regions of time where systematic and statistical uncertainties are largest.  The idea was suggested in Ref.~\cite{Lehner:2017kuc} and first applied in Ref.~\cite{RBC:2018dos}. The same time-window must then of course be applied to the $R_{e^+e^-}$ data. The BMW collaboration~\cite{Borsanyi:2020mff} did this for a time-window between 0.4 and 1\,fm and obtained a lattice result that is higher by 3.7$\sigma$ than the corresponding data-driven value. Although not a 5$\sigma$ discrepancy, this is more significant than the tension seen between their lattice result and the data-driven value for the full HVP contribution. The time-window used was sufficient to capture about one third of the HVP. The difference between the lattice and data-driven results is $7.0(1.9)\times10^{-10}$, corresponding to about one quarter of the tension between the Muon g-2 experimental result and the SM value for $a_{\mu}$ using data-driven values for the HVP. Recent independent lattice determinations of the HVP in this window by the Mainz/CLS collaboration~\cite{Ce:2022kxy} and by the ETM Collaboration~\cite{Alexandrou:2022amy} using different lattice QCD actions both give results in good agreement with that of BMW and with a similar uncertainty. For other lattice results from this time window see Refs.~\cite{RBC:2018dos, Aubin:2019usy, Lehner:2020crt, Giusti:2021dvd, Wang:2022lkq, Aubin:2022hgm}. 

Here we study whether wider windows can be used to sharpen the comparison between data-driven and lattice results by capturing a larger fraction of the total HVP without increasing the lattice-QCD uncertainties. This approach will help establish whether or not there is a significant disagreement between data-driven evaluations and lattice-QCD calculations of the HVP. This question goes to the heart of the interpretation of the tension between experiment and the current SM expectation for $a_{\mu}$ as a signal for new physics.

To demonstrate the advantage of wider windows, we compare results obtained from the simulations described in the Fermilab Lattice/HPQCD/MILC 2019~analysis of the HVP~\cite{FermilabLattice:2019ugu} (along with those in Refs.~\cite{Chakraborty:2014mwa,Hatton:2020qhk,FermilabLattice:2021hzx}) with results obtained from the KNT19 analysis~\cite{Keshavarzi:2019abf} of $R_{e^+e^-}$ data~\cite{Bai:1999pk,Akhmetshin:2000ca,Akhmetshin:2000wv,Achasov:2000am,Bai:2001ct,Achasov:2002ud,Akhmetshin:2003zn,Aubert:2004kj,Aubert:2005eg,Aubert:2005cb,Aubert:2006jq,Aulchenko:2006na,Achasov:2006vp,Akhmetshin:2006wh,Akhmetshin:2006bx,Akhmetshin:2006sc,Aubert:2007ur,Aubert:2007ef,Aubert:2007uf,Aubert:2007ym,Akhmetshin:2008gz,Ambrosino:2008aa,Ablikim:2009ad,Aubert:2009ad,Ambrosino:2010bv,Lees:2011zi,Lees:2012cr,Lees:2012cj,Babusci:2012rp,Akhmetshin:2013xc,Lees:2013ebn,Lees:2013uta,Lees:2014xsh,Achasov:2014ncd,Aulchenko:2014vkn,Akhmetshin:2015ifg,Ablikim:2015orh,Shemyakin:2015cba,Anashin:2015woa,Achasov:2016bfr,Achasov:2016lbc,TheBaBar:2017aph,CMD-3:2017tgb,TheBaBar:2017vzo,Kozyrev:2017agm,Anastasi:2017eio,Achasov:2017vaq,Xiao:2017dqv,TheBaBar:2018vvb,Anashin:2018vdo,Achasov:2018ujw,Lees:2018dnv,CMD-3:2019ufp}.
The layout of the paper is as follows. In Sec.~\ref{sec:lattice} we describe the lattice calculation and the imposition of a time-window on the integral. Sections~\ref{sec:connll} and~\ref{sec:connsc} discuss the quark-line connected contributions, Sec.~\ref{sec:discwindow} covers the disconnected contribution using results from an ongoing blinded analysis (with further details in Appendix~\ref{sec:disc}), and Sec.~\ref{sec:QEDSIB} discusses the uncertainty we allow for missing QED and strong-isospin breaking contributions. Section~\ref{sec:R} describes the calculation of the HVP from $R_{e+e^-}$ data imposing the same time-window, while Sec.~\ref{sec:results} compares the two. We draw conclusions and present an outlook in Sec.~\ref{sec:conclusions}. 

\section{The HVP from lattice QCD with a time-window}
\label{sec:lattice}

\begin{table*}[tbh]
    \caption{Parameters of the MILC HISQ $n_f=2+1+1$ QCD gauge-field ensembles~\cite{MILC:2012znn}.  The first column labels the ensembles, the second shows the approximate lattice spacing, while the third, fourth and fifth list the bare lattice up/down (set equal and denoted $m_l$) , strange, and charm sea-quark masses in lattice units.  The sixth column gives the ratio of the gradient-flow scale $w_0$~\cite{Borsanyi:2012zs} to the lattice spacing; to convert quantities in lattice-spacing units to\,GeV we use $w_0=0.1715(9)$~fm~\cite{Dowdall:2013rya}.    The seventh column lists the taste-Goldstone sea-pion masses; these were obtained from fits of pseudoscalar-current two-point correlators as in Ref.~\cite{MILC:2012znn}.  The eighth column gives the lattice volumes. The final two columns give the number of configurations analyzed and the number of random-wall time sources used per configuration. \vspace{1mm}}
    \label{tab:ensembles}
\begin{ruledtabular}
\begin{tabular}{llccccccccccc}
Set & $\approx a$ (fm) & $am_l^{\rm sea}$ & $am_s^{\rm sea}$ & $am_c^{\rm sea}$ & $w_0/a$ &  $M_{\pi_5}$ (MeV) & $(L/a)^3 \times (T/a)$ & $N_{\rm conf.}$  & $N_{\rm wall}$ \\
\hline
1 & 0.15 & 0.00235 & 0.0647 & 0.831 & 1.13670(50) &  133.04(70) &  $32^3 \times 48$ & 997 & 16 \\
2 & 0.12 & 0.00184 & 0.0507 & 0.628 & 1.41490(60) &  132.73(70) &  $48^3 \times 64$ & 998 & 16 \\
3 & 0.09 & 0.00120 & 0.0363 & 0.432 & 1.95180(70) &  128.34(68) &  $64^3 \times 96$ & 1557 & 16 \\
4 & 0.06 & 0.0008 & 0.022 & 0.260 & 3.0170(23) &  134.95(72) &  $96^3 \times 192$ & 1170 & 16 \\
\end{tabular}

\end{ruledtabular}
\end{table*}

Lattice-QCD calculations of the HVP proceed by calculating Euclidean time vector-vector correlation functions on sets of gluon field configurations:
\begin{equation}
\label{eq:corr}
G_{ff^\prime}(t) = Q_f Q_{f^\prime} \sum_{\vec{x}}Z^2_V \langle j^i_f (\vec{x},t)j^i_{f^\prime}(0)\rangle \, .
\end{equation} 
Here $f$ and $f^\prime$ are flavour indices, $Q_f$ is the electric charge for that flavour in units of $e$, $i$ is a spatial index and $j^i=\overline{q}\gamma^iq$. $Z_V$ is the renormalisation factor for the lattice vector (electromagnetic) current needed to match it to that in continuum QCD. The dominant quark-line connected correlators (upper picture in Fig.~\ref{fig:bubble}) are diagonal in flavour while the quark-line disconnected correlators (lower picture) are not. 

The contribution to $a_{\mu}$ from $G_{ff^\prime}(t)$ is then given by an integral over time~\cite{Bernecker:2011gh}: 
\begin{equation}
\label{eq:int}
a_{\mu,ff^{\prime}}^{\mathrm{HVP}}=\left(\frac{\alpha}{\pi}\right)^2 \int_0^{\infty} \! dt\, G_{ff^\prime}(t) K_G(t) \, ,
\end{equation}
where the kernel $K_G(t)$ vanishes at $t=0$. $K_G(t)$ grows with increasing $t$ as $G(t)$ falls exponentially, so the integrand of Eq.~\ref{eq:int} peaks at a value of $t$ that is determined by the masses of the vector states present in $G(t)$. 

Lehner~\cite{Lehner:2017kuc} suggested a windowing approach for integrating lattice-QCD results over $t$ (see also Ref.~\cite{Bernecker:2011gh}). This consists of multiplying the integrand of Eq.~\eqref{eq:int} by a difference of two step functions to  integrate over a time-region between $t=t_0$ and $t=t_1$ only (an `intermediate window'), softening the edges of the time region with a time-width $\Delta t$. The idea is to cut out large $t$ values from the lattice-QCD integral to reduce statistical and systematic uncertainties that grow at large $t$. The reason given for the lower $t$ limit, $t_0$, was to reduce discretisation errors from lattice QCD, although the kernel function suppresses small-$t$ contributions. Parameter values $t_0=$ 0.4\,fm, $t_1 =$ 1.0\,fm and $\Delta t = $ 0.15\,fm were suggested in Ref.~\cite{RBC:2018dos} as corresponding to the region where data-driven results have largest relative uncertainty and so there is potential for lattice-QCD results to complement them. The BMW collaboration adopted this time-window for their analysis in Ref.~\cite{Borsanyi:2020mff}. 

Here we use a simpler, and larger, time-window that nevertheless shares the important property that statistical and systematic uncertainties from lattice QCD are much reduced. Our window simply adapts the intermediate window of Ref.~\cite{Lehner:2017kuc} to drop the lower time parameter $t_0$.\footnote{A similar window is used in Ref.~\cite{Lehner:2017kuc} but only for very short time distances.} Our one-sided window then extends from $t=0$ upwards to $t_1$ with a rounded edge of width $\Delta t$. The window function that multiplies the integrand of Eq.~\eqref{eq:int} is given by 
\begin{equation}
\label{eq:window}
\Theta(t,t_1,\Delta t) = \frac{1}{2} \left[1-\tanh\left(\frac{t-t_1}{\Delta t}\right)\right] \, .
\end{equation} 
The contribution to $a_{\mu}$ from this window is then 
\begin{equation}
\label{eq:intwin}
a_{\mu}^w(t_1,\Delta t)=\left(\frac{\alpha}{\pi}\right)^2 \int_0^{\infty} \!dt\, G_{ff^\prime}(t) K_G^w(t)\, ,
\end{equation}
with a modified kernel,
\begin{equation}
    K_G^w(t) \equiv K_G(t) \Theta(t,t_1,\Delta t).
\end{equation}
This time window, for suitable values of $t_1$, can provide a good basis for a stringent comparison of lattice and data-driven results. as we show below. 

In what follows we examine the contributions coming from different flavours of quark.

\subsection{Connected iso-symmetric contributions from $u/d$ quarks}
\label{sec:connll}

 \begin{figure}
    \parbox[center]{0.92\linewidth}{
    \begin{flushright}   
    \includegraphics[scale=0.9]{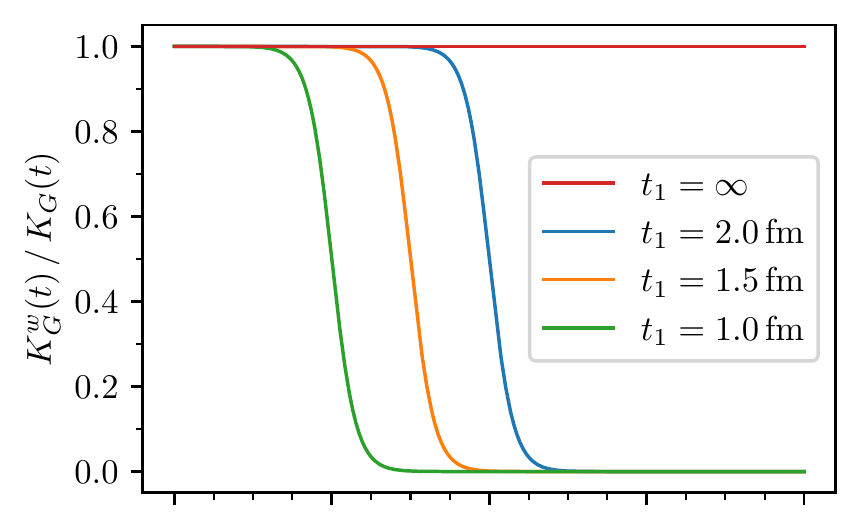}\!\! \\
    \vspace{-2ex}
    \includegraphics[scale=0.9]{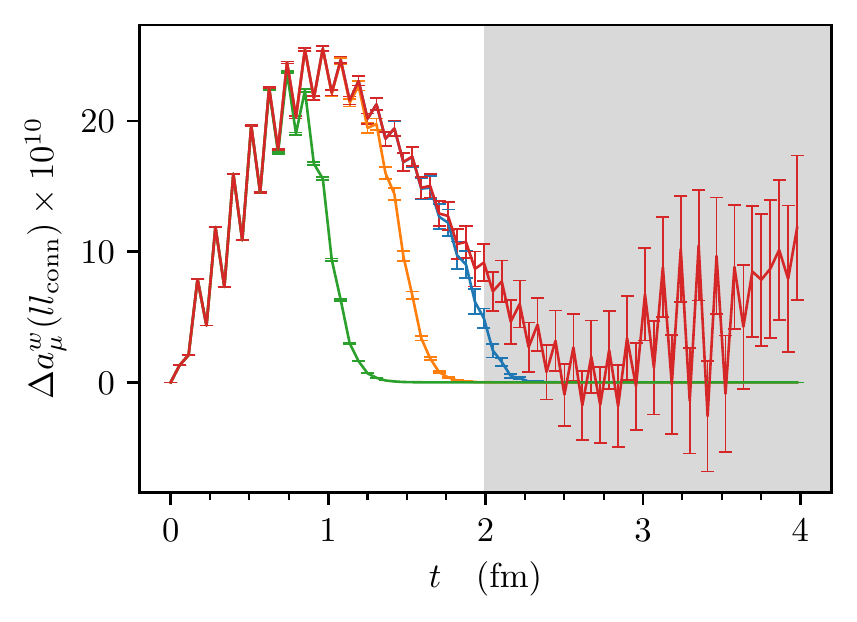}
    \end{flushright}
    }\vspace{-3ex}
    \caption{Top: Ratio of kernels $K_G^w/K_G=\Theta$ from Eq.~\eqref{eq:window}
   as a function of $t$ with one-sided windows where (upper curve to lower curve)
   $t_1=\infty$ (red), $t_1=2.0$ (blue), 1.5 (orange), and 1.0 (green)\,fm and $\Delta t =0.15$\,fm. 
    Bottom: Integrand $\Delta a_\mu^w$ of Eq.~\ref{eq:intwin} from the lattice $ll$ connected correlator
    $G(t)$ on the $a =0.06$\,fm lattices
    for each $t$ on the lattice out to 4\,fm; we have insufficient statistics to give reliable results 
    for $t>2$\,fm (grey shading). Results are shown for
    the one-sided windows in the top pane with corresponding colours. The one-sided window cuts out the less useful correlator results from the integrand. The oscillations in the correlator are a consequence of using staggered quarks.  }
    \label{fig:win-t}
\end{figure}

\begin{table*}
    \caption{\label{tab:results}Lattice contributions to $a_\mu^w$ with one-sided windows of
        varying time extent~$t_1$ and rounding width, $\Delta t$ of 0.15\,fm (Eq.~\eqref{eq:window}). All values are in units of $10^{-10}$. Contributions are given from the connected light-quark vacuum polarisation 
        ($ll_\mathrm{conn}$), the light and strange quark disconnected 
        vacuum  polarisation ($(ll+ss)_\mathrm{disc}$), and from (connected) contributions from 
        the $s$, $c$ and $b$ quarks. 
        The fraction of the HVP (computed using $R_{e^+e^-}$ data) that is included in the window is listed under~\%\,HVP.
        The sum of these contributions
        $a_\mu^w(\mathrm{latt})$ can be compared with results from 
        $R_{e^+e^-}$; the difference is listed in the 
        last column. The second error on $a_\mu^w(\mathrm{latt})$ accounts for corrections from 
        QED and strong isospin breaking.}
        \begin{ruledtabular}\begin{tabular}{clllllllll}
            $t_1$ & $ll_\mathrm{conn}$ & $(ll+ss)_\mathrm{disc}$ & $s$ & $c$ & $b$ & \%\,HVP & $a_\mu^w(\mathrm{latt})$ & $a_\mu^w(R)$ & $a_\mu^w(\mathrm{latt})-a_\mu^w(R)$ \\
            \hline
            0.5 & 72.9 (2) & $-$0.02 (0) & 13.5 (1) & 13.20 (5) & 0.30 (1) & 14.2\% & 99.9 (2)(2) & 98.3 (6) & 1.6 (7) \\
            1.0 & 253.7 (8) & $-$0.98 (22) & 36.3 (2) & 14.63 (5) & 0.30 (2) & 43.0\% & 304.0 (9)(6) & 297.9 (1.2) & 6.0 (1.6) \\
            1.5 & 436.0 (3.2) & $-$4.34 (1.00) & 48.9 (3) & 14.64 (5) & 0.30 (2) & 70.0\% & 495.5 (3.4)(1.0) & 485.0 (1.7) & 10.4 (3.9) \\
            2.0 & 546.5 (7.3) & $-$8.98 (2.14) & 52.5 (3) & 14.64 (5) & 0.30 (2) & 86.1\% & 605.0 (7.6)(1.2) & 596.3 (2.1) & 8.7 (8.0) \\
            3.0 & 610.6 (18.5) & $-$13.83 (3.40) & 53.4 (4) & 14.64 (5) & 0.30 (2) & 97.2\% & 665.1 (18.8)(1.3) & 673.1 (2.3) & $-$8.0 (19.0) \\
        \end{tabular}\end{ruledtabular}
\end{table*}

\begin{figure}
    \includegraphics[scale=0.9]{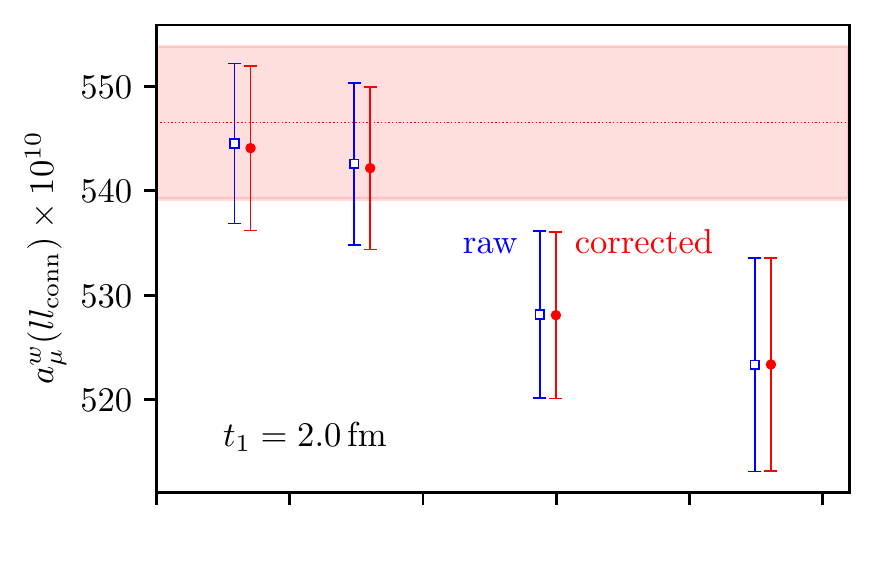}\\
    \vspace{-4ex}
    \includegraphics[scale=0.9]{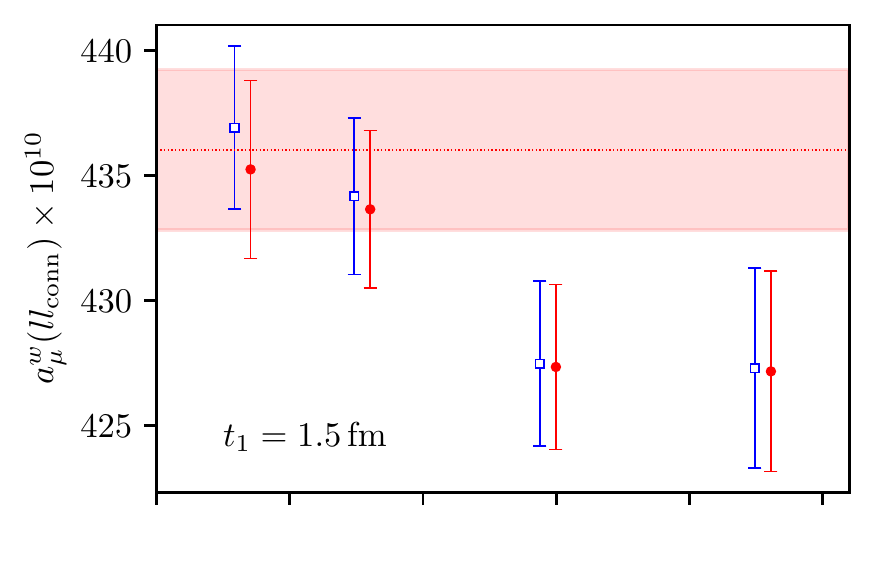}\\
    \vspace{-4ex}
    \includegraphics[scale=0.9]{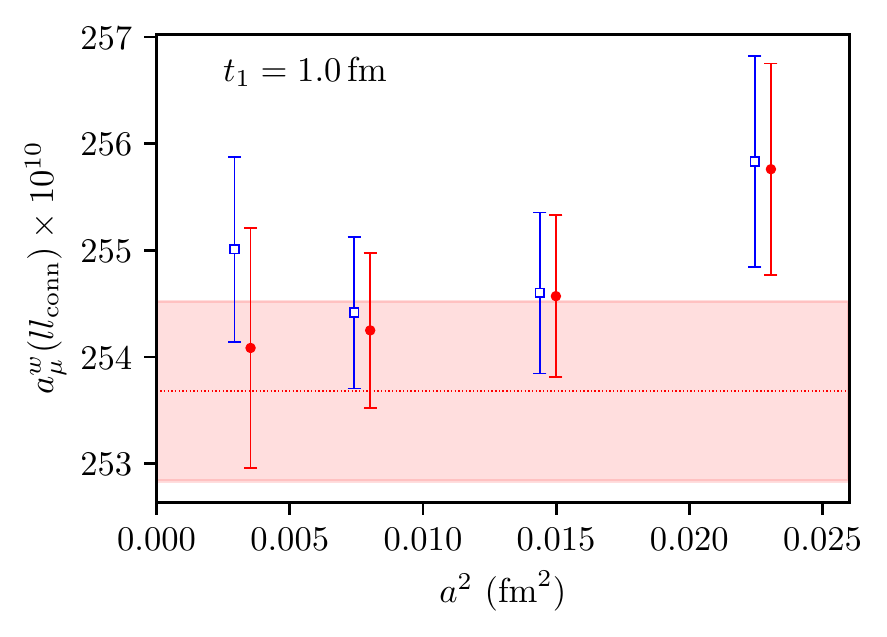}\\
    \vspace{-3ex}
    \caption{Lattice-QCD results for windowed HVP contributions from $ll$ connected correlation functions for time-windows 
    having $t_1=2.0$, 1.5, and 1.0\,fm and $\Delta t = 0.15$\,fm plotted versus the lattice 
    spacing squared. The filled red circles are corrected for 
    finite-volume errors and small mistunings
    of the light-quark mass. The open blue squares are the 
    uncorrected results. The red band is the result obtained by 
    extrapolating the corrected results to $a^2=0$.}
    \label{fig:t1vsa}
\end{figure}

\begin{figure}
    \includegraphics[scale=0.9]{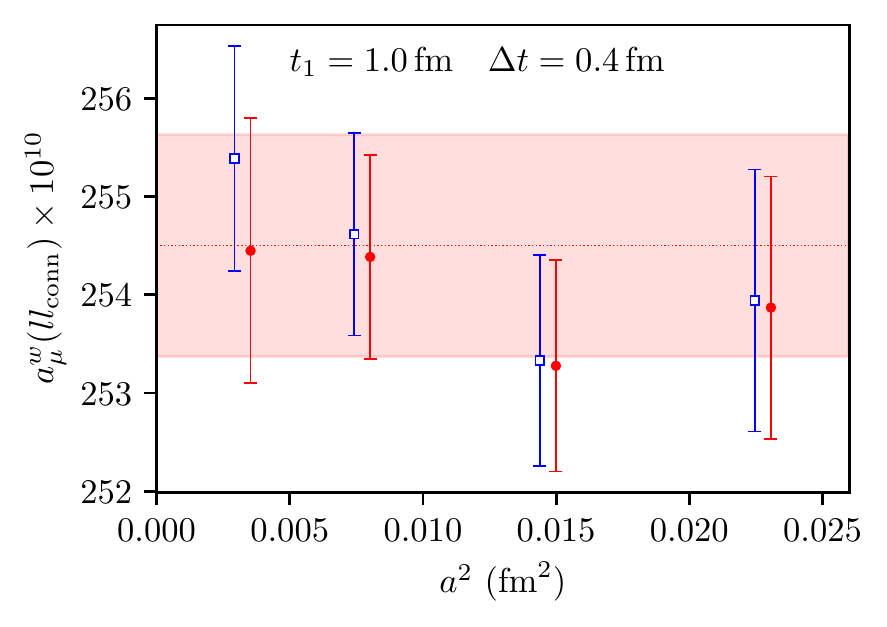}
    \caption{Same as the bottom pane of Figure~\ref{fig:t1vsa} but with a
 wider window edge of $\Delta t=0.4$\,fm. Discretization errors are 
    significantly smaller for the corrected results (red circles) with the larger~$\Delta t$.}
    \label{fig:big-dt}
\end{figure}

By far the largest contributions to the HVP come from connected correlators involving $u$ or $d$ currents. In this section we discuss isospin symmetric contributions; corrections from strong-isospin breaking and QED are discussed below, in Sec.~\ref{sec:QEDSIB}. We therefore take $u$ and $d$ quarks to have the same mass and call them $l$ (for light) quarks. We use the correlation functions that were previously used to determine the full HVP contribution in Ref.~\cite{FermilabLattice:2019ugu}. 

The calculations use the HISQ action~\cite{Follana:2006rc} for the valence quarks on gluon field configurations that include $n_f=2+1+1$ flavours of HISQ sea quarks generated by the MILC collaboration~\cite{MILC:2010pul,MILC:2012znn}. An advantage of the HISQ action is its very good control of discretisation effects because $a^2$ errors are removed at tree-level. We will demonstrate the impact of lattice spacing effects for our time-windowed results below. 

The $ll$ connected correlators were calculated for a mass~$m_l$ tuned to the physical average of the $u$ and $d$ quarks' masses  
on gluon field configurations with multiple values of the lattice spacing, $a$, covering the range from $a =$ 0.15\,fm down to $a =$ 0.06\,fm~\cite{FermilabLattice:2019ugu}. Here we use the same correlators except for the high-statistics set for $a=0.15$\,fm, which we omit because it is part of a larger (blinded) study that is in progress. Extensive analysis of these correlators was undertaken in 2019; see Ref.~\cite{FermilabLattice:2019ugu} for more details.\footnote{We also dropped a small number of defective correlator measurements which we discovered in the 0.06\,fm~data set.} The parameters of the ensembles we use for the $ll$ calculation are given in Table~\ref{tab:ensembles}.

Among the HVP contributions from connected correlators, the $ll$ contribution is the most difficult to calculate well on the lattice. This is primarily because of the rapid growth of statistical noise with increasing time separation~$t$ between the two vector currents in the correlator: the mass parameter ($2m_{\pi}$) that controls the exponential fall-off of the noise is much smaller than that which controls the exponential fall-off of the signal (this mass being $m_{\rho}$ over the time interval that dominates~$a_\mu^\mathrm{HVP}$)~\cite{Parisi:1983ae, Lepage:1989hd}. 

In Ref.~\cite{FermilabLattice:2019ugu} contributions from $t$ values larger than 2\,fm rapidly became unreliable given the statistics used there (see Fig. 2 in that paper). The problem was addressed by replacing Monte Carlo data for the correlator at large times $t>t^*$ with a correlator extrapolated from fits to Monte Carlo data dominated by the more precise results for $t<t^*$. Here this will not be necessary because we will choose $t$-windows that exclude most of the region~$t>t^*=2$\,fm.

The top pane of Fig.~\ref{fig:win-t} shows the window function of Eq.~\eqref{eq:window} for three values of $t_1$: 1.0, 1.5 and 2.0\,fm (as well as $t_1=\infty$). The lower pane shows the integrand of Eq.~\eqref{eq:intwin} including these window functions for connected $ff^{\prime}=ll$ correlator results at our finest lattice spacing, $a =$ 0.06\,fm. Note how the window functions cut out the lattice results with large statistical errors, from $t>2$\,fm.

A second issue for simulations (like ours) using staggered quarks is the $a^2$~errors caused by mass splittings between pions of different taste~\cite{Chakraborty:2016mwy}. These errors were as large as~11\% in Ref.~\cite{FermilabLattice:2019ugu}, which used the chiral model of Ref.~\cite{Chakraborty:2016mwy} to remove them. Again this correction is not needed here because the effect comes primarily from large values of $t$ that are excluded by our windows; any residual $a^2$ dependence is much smaller and can be extrapolated away.

We have calculated the connected $ll$ contribution to $a_\mu^w$ for a variety of $t_1$~values from 0.5\,fm to 3.0\,fm using the correlators from Ref.~\cite{FermilabLattice:2019ugu}. For each $t_1$, we calculate $a_\mu^w$ for each of our four lattice spacings. Adapting the procedure outlined in Ref.~\cite{FermilabLattice:2019ugu}, we correct each of the results to remove systematic errors due to the finite volume of the lattice and (small) mistunings of the light-quark mass (but not those due to the pion taste splittings). These corrections are specified as functions of the energy flowing through the correlator in Refs.~\cite{Chakraborty:2016mwy,FermilabLattice:2019ugu}; we Fourier transform them to $t$~space so we can apply the windows. The corrections are less than~1\% for the values with~$t_1$ between~1 and~2\,fm.\footnote{We assign an uncertainty to the finite-volume correction that is 10\% of the $\pi\pi$~contribution (first term in Eq.~(B33) of Ref.~\cite{Chakraborty:2016mwy}) or 30\% of the $\rho$ contribution (second term in Eq.~(B33)), whichever is larger. The $\pi\pi$~contribution is the lowest-order contribution in our chiral model, while the $\rho$~contribution enters at the next order (see Fig. 8 in~\cite{Chakraborty:2016mwy}). The $\rho$ contribution is only 15\% of the $\pi\pi$ contribution absent windows, but it is less affected by the windows and so becomes more competitive for smaller~$t_1$s (and dominant for $1 \le t_1 \le 2$\,fm). We parameterise the corrections with (5,5) Pad\'{e} approximants.}

Sample results for $a_\mu^w$ with $t_1$ between~1 and 2\,fm are shown versus~$a^2$ in Fig.~\ref{fig:t1vsa}, where both corrected and uncorrected results are plotted. We extrapolate the corrected $a_\mu^w$ results to zero lattice spacing using the fitting procedure and priors described in Ref.~\cite{FermilabLattice:2019ugu}, except that we quadruple the width of the priors associated with $a^2$ and $a^4$ errors to account for larger discretization errors (since we are not correcting for taste splittings). The extrapolated results are shown as dotted red lines in Fig.~\ref{fig:t1vsa}. 

The corrected results in Fig.~\ref{fig:t1vsa} show $a^2$ errors at the largest lattice spacing of  $-4.2$\%, $-2.0$\%, and $0.8$\% as $t_1$ decreases from~2\,fm to~1\,fm. These are substantially smaller than the error for the total HVP (i.e. without a window and uncorrected for taste-splitting), as expected. The error changes sign for the smallest~$t_1$ because of $a^2$~effects caused by the edge of the window at~$t_1$ (recall that the transition width $\Delta t$ equals our coarsest lattice spacing). This edge has little effect when $t_1$ is large (see the bottom pane of Fig.~\ref{fig:win-t}) because there is little contribution to $a_\mu^w$ coming from the edge region. For $t_1<1$\,fm, however, the edge sits in the most important region contributing to $a_\mu^w$ and therefore causes substantial $a^2$ errors. Increasing $\Delta t$ decreases the $a^2$~errors caused by the window, as is evident by comparing the bottom pane of Fig.~\ref{fig:t1vsa} ($t_1=1$\,fm, $\Delta t=0.15$\,fm) with Fig.~\ref{fig:big-dt} ($t_1=1$\,fm, $\Delta t=0.4$\,fm).
 
Table~\ref{tab:results} lists windowed HVP results from the $ll$ connected correlators for several values of $t_1$ from $0.5$\,fm\footnote{To get a good continuum limit fit, the coarsest lattice spacing must be dropped for $t_1=0.5$\,fm.} to $3.0$\,fm.
 In what follows, we will concentrate on three $t_1$ values\,---\,1\,fm, 1.5\,fm and 2\,fm\,---\,to avoid large $a^2$~errors (from the window) at small~$t_1$ and large statistical errors at large~$t_1$. The windows with $t_1 = 1.5$\,fm and 2\,fm are the most useful because they capture large fractions of the total HVP: 70\% and 86\%, respectively.

\subsection{Connected contributions from $s$, $c$, and $b$ quarks}
\label{sec:connsc}

\begin{figure}
    \parbox[center]{0.92\linewidth}{
    \begin{flushright}   
       \includegraphics[scale=0.9]{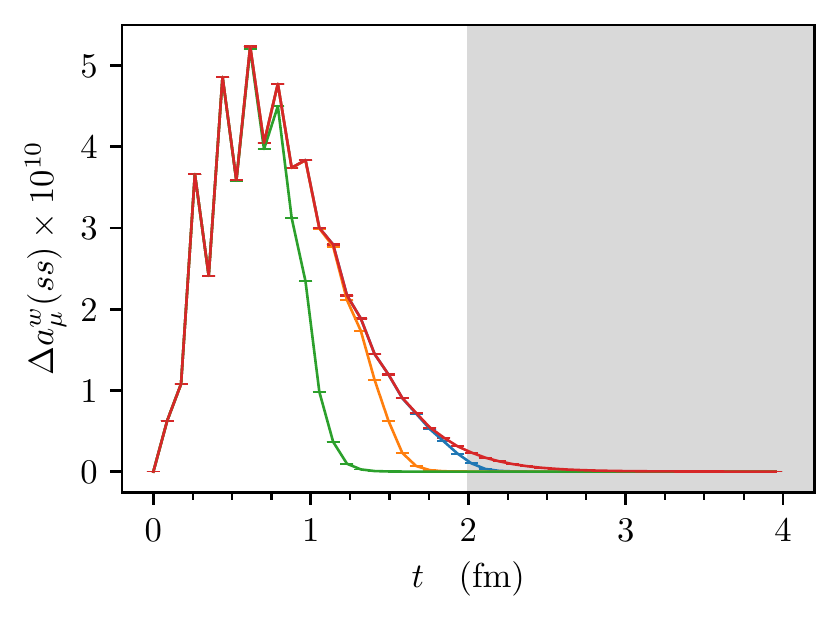}
    \end{flushright}
    }\vspace{-3ex}
    \caption{ Contributions to $a_\mu^w$ from lattice $ss$ connected correlator
    $G(t)$ on the $a =0.09$\,fm lattices
    vs. $t$ out to 4\,fm. Results
    from $t>2$\,fm are greyed out to indicate where the $ll$ results develop large uncertainties; there is no such issue with the $ss$ case. Results are shown for
    the one-sided time-windows where (from top to bottom) $t_1=\infty$ (red),
    2.0 (blue), 1.5 (orange), and 0.5 (green)\,fm. Note that the contributions peak earlier in $t$ than for the $ll$ case and are almost entirely included in the $t_1=2$\,fm window. The oscillations in the correlator are a consequence of using staggered quarks.  }
    \label{fig:swin}
\end{figure}

The contribution to $a_\mu^\mathrm{HVP}$ from connected $s$-quark correlators is much easier to calculate than that from $u$ and $d$~quarks, because the statistical noise at large~$t$ is proportionately much smaller (because of the smaller difference between masses controlling signal and noise). Using the $ss$ correlators described in~\cite{Chakraborty:2014mwa}, Fig.~\ref{fig:swin} shows the effect of the one-sided window on the $ss$ case, comparable to that for the $ll$ case in the lower plot of Fig.~\ref{fig:win-t}. 
Because the $ss$ correlator falls more rapidly with $t$ than in the $ll$ case, being controlled by the larger mass of the $\phi$ meson, the contributions peak at lower $t$ than in Fig.~\ref{fig:win-t}, and a higher proportion of the full contribution is captured by our windows. Finite-volume errors are negligible for our volumes for the $s$ quark case~\cite{Chakraborty:2014mwa}. 

We calculate the effect of windows on the $ss$ correlators of Ref.~\cite{Chakraborty:2014mwa} using Eq.~(\ref{eq:intwin}). We use the same fitting procedure as in that paper to extrapolate to $a^2=0$, except that we have corrected an error that led us to overestimate the lattice spacing uncertainty there. There are two anti-correlated sources of lattice spacing uncertainty: one a direct effect on the HVP contribution and the other from the tuning of the $s$-quark mass. When both are included correctly, the contribution to the total $ss$ HVP error budget from the lattice spacing uncertainty (from $w_0$ and $w_0/a$) falls from 1.1\% quoted in Ref.~\cite{Chakraborty:2014mwa} to 0.37\%. Since this is the dominant uncertainty, this also causes a reduction of the total uncertainty on the full connected $s$ quark HVP contribution, giving $a_{\mu}^s=53.41(35)\times 10^{-10}$ (with (35) replacing the (59) quoted in Ref.~\cite{Chakraborty:2014mwa}). This result supersedes that in Ref.~\cite{Chakraborty:2014mwa}. 

Table~\ref{tab:results} then lists the extrapolated results for $a_{\mu}^w$ for the $ss$ case for a range of $t_1$ values. For the $t_1=0.5$\,fm window we again drop the $a=0.15$\,fm lattices from the $a^2$ extrapolation because of large discretisation errors. A higher proportion of the $s$-quark contribution is included in $a_{\mu}^w$ than for the $ll$ case -- 91\% for the 1.5\,fm window and 98\% for the 2\,fm window --
 in agreement with what is seen in Fig.~\ref{fig:swin}. 

Contributions to $a_{\mu}$ from $c$ and $b$~quarks are much smaller. For the contribution from $c$ quarks here we use the recent results determined by the HPQCD collaboration that include also the effect of QED for the valence quarks~\cite{Hatton:2020qhk}. For the (negligible) $b$ quark contribution we use HPQCD results from Ref.~\cite{Hatton:2021dvg}. The $c$ and $b$ correlators are precisely calculated in lattice QCD and have previously been compared with results from $R_{e^+e^-}$ for each quark (by subtraction of the contributions from other flavours using high-order QCD perturbation theory) at the level of correlator moments as well as for the HVP contribution~\cite{Donald:2012ga, Colquhoun:2014ica, Nakayama:2016atf, Hatton:2020qhk, Hatton:2021dvg}. 

To implement windows for $c$ and $b$ correlators, we first construct a $(2,2)$ Pad\'e approximant for the subtracted vacuum polarization function~$\hat\Pi(q^2)$ from moments of the correlators (extrapolated to $a^2=0$)~\cite{Chakraborty:2014mwa}. We then Fourier transform this function to obtain a (Euclidean) correlator~$G(t)$ from which we can calculate~$a_\mu^w$ using Eq.~(\ref{eq:intwin}). The results are  again listed in Table~\ref{tab:results} for a range of $t_1$ values. Because the $c$ and $b$ correlators have such strong exponential decay with $t$, essentially all of their HVP contribution is contained in even the shortest time-window that we consider ($t_1 =$0.5\,fm). 

\subsection{Disconnected contributions}
\label{sec:discwindow}

\begin{figure}
    \includegraphics[scale=0.9]{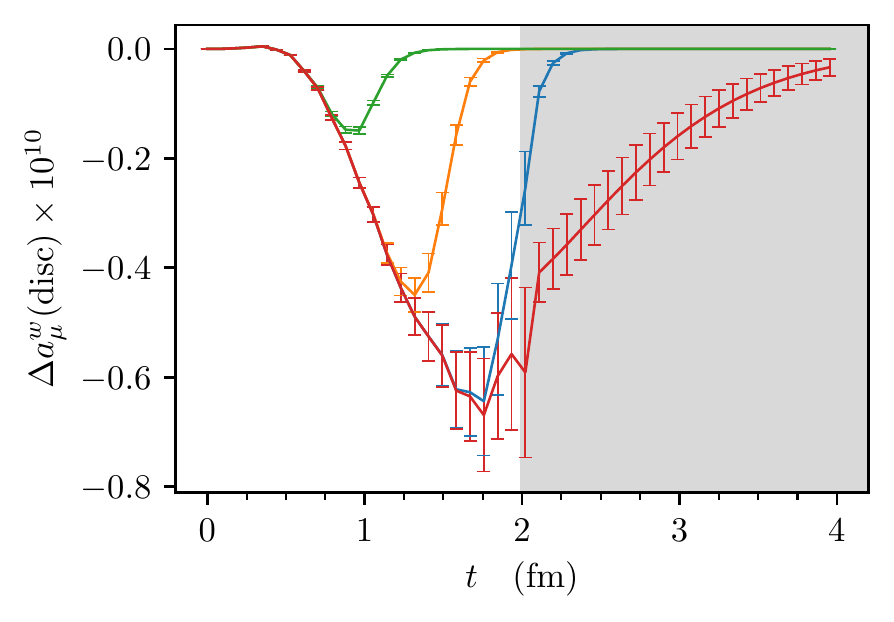}
    \caption{Contributions to $a_\mu^w$ from combined quark-line disconnected lattice correlators
    for $l$ and $s$  quarks on the $a =0.09$\,fm lattices
    for each $t$ on the lattice out to 4\,fm; we have insufficient statistics to give reliable results
    for $t>2$\,fm (grey shading). Results are shown for
   our one-sided windows with (from bottom to top) $t_1=\infty$ (red),
    2.0 (blue), 1.5 (orange), and 0.5\,fm (green). The time-windows cut out the less reliable correlator results from the integrand. }
    \label{fig:Gdisc}
\end{figure}

The quark-line disconnected contribution to the HVP is small because it is suppressed by quark mass differences~\cite{Blum:2002ii}. 
It is large enough, however, that a moderately accurate calculation is needed and this is quite challenging. Appendix~\ref{sec:disc} provides technical details of our ongoing calculation of this contribution~\cite{FermilabLattice:2021hzx}, which is currently blinded. 

Because of the blinding, we do not show explicit results here for either the full or windowed contributions to $a_{\mu}$ from the disconnected correlators of Ref.~\cite{FermilabLattice:2021hzx}. 
Instead, as discussed in Appendix~\ref{sec:disc}, we determine the ratio $a_{\mu}^w/a_{\mu}$ from these correlators on the 0.09\,fm configurations for each $t_1$ value in Table~\ref{tab:results}. The blinding factor cancels in the ratios. Next we correct the ratios $a_{\mu}^w/a_{\mu}$ for finite-volume effects and mistunings of the light quarks, taking a 10\% uncertainty on the corrections (which are all smaller than~$0.4\times10^{-10}$).\footnote{We neglect the next-to-leading order corrections here because contributions from the $\rho$ and $\omega$ tend to cancel~\cite{Chakraborty:2015ugp}.} 
We also take an additional overall uncertainty of~15\% in the ratio to account for residual $a^2$ errors on the 0.09\,fm~lattice.\footnote{The $a^2$~error on the disconnected~$a_\mu$ without a window is about 30\% for the 0.09\,fm~lattice. We take a smaller error of~15\% for the ratio since this error tends to cancel in the ratio for large values of~$t_1$. It is likely an overestimate for $t_1\ge1.5$\,fm. It is possibly an underestimate for smaller $t_1$s, but negligible compared to other errors contributing to the total windowed HVP.}
Finally we multiply these ratios by the BMW result for the disconnected contribution, adjusted to infinite volume (using 1/9 of the finite-volume correction quoted for the full HVP)~\cite{Borsanyi:2020mff}: $-15.46(1.82)\times 10^{-10}$. This procedure yields the results listed in column~3 of Table~\ref{tab:results}. The procedure exaggerates the uncertainties coming from the disconnected contribution; an unblinded analysis of our data, with multiple lattice spacings, would have significantly smaller errors.

We choose to employ the BMW determination~\cite{Borsanyi:2020mff} here because it has the smallest uncertainty from lattice QCD to date. Earlier lattice-QCD results for this quantity, while generally more uncertain, are consistent with the value~\cite{Budapest-Marseille-Wuppertal:2017okr, RBC:2018dos, Gerardin:2019rua}. Further lattice-QCD results of comparable quality to that of BMW are needed. We note also a recent data-driven determination of the sum of $s$-quark connected and disconnected contributions using $R_{e^+e^-}$ results along with the $\tau \rightarrow K^-K^0\nu_{\tau}$ distribution~\cite{Boito:2022rkw}. Subtracting the lattice average for the connected $s$-quark contribution gives a value~\cite{Boito:2022rkw} for the disconnected contribution of $-13.3(1.5)\times 10^{-10}$ using the KNT19 $R_{e^+e^-}$ compilation~\cite{Keshavarzi:2019abf} and $-14.6(2.1)\times 10^{-10}$ using the DHMZ compilation~\cite{Davier:2019can}. These results agree within uncertainties with the BMW result, as would be expected in a scenario with no new physics, and have comparable errors. 

The values shown in column~3 of Table~\ref{tab:results} include contributions from the quark-line disconnected contributions for $l$ and $s$ quarks combined, calculated using Eqs.~\eqref{eq:int} and~\eqref{eq:intwin}. The uncertainty includes both that from the windowed ratio calculated from the correlators of Ref.~\cite{FermilabLattice:2021hzx} and that from the BMW result used for normalisation combined in quadrature. Note that the relative importance of these contributions decreases with decreasing~$t_1$ from 2\% of the total $a_{\mu}^w(\mathrm{latt})$ for $t_1=3.0$\,fm to less than 1\% with~$t_1=1.5$\,fm. 

The rapid falloff in the  disconnected contribution with decreasing~$t_1$ can be understood by examining the contribution to~$a_\mu^w$ coming from the disconnected correlator~$G^{\mathrm{disc}}(t)$ for each value of~$t$ separately, as plotted in Fig.~\ref{fig:Gdisc}.  The contribution to $a_{\mu}$ from $G^{\mathrm{disc}}(t)$ peaks at a larger value of $t$ than is the case for the $ll$ connected correlation function (Fig.~\ref{fig:win-t}) since the contributions from the isospin-1 ($\rho$) and isospin-0 ($\omega$, $\phi$) states that dominate the connected contributions largely cancel in the disconnected case~\cite{Chakraborty:2015ugp}. Because of this more `infra-red' behaviour, the one-sided time-windows greatly suppress the disconnected contribution to $a_{\mu}^w$ and successfully cut out the less reliable results at larger $t$ values. This means that a precise value for $a_{\mu}^w$ from the disconnected correlation function can readily be determined. Our uncertainty here is a result of the blinding (see Appendix~\ref{sec:disc}), and will improve significantly once that is removed. This limitation will not have a large impact on the uncertainty of the total $a_{\mu}^w(\mathrm{latt})$ however.

\subsection{QED and strong-isospin breaking effects} 
\label{sec:QEDSIB} 

In the real world, the up and down quarks differ in mass and electric charge, giving rise to small (sub-percent) corrections to a lattice calculation done in pure QCD in the isospin-symmetric limit, as done here for the $ll$ and $ss$ contributions. These additional effects can change the value of $a_{\mu}$ directly but also indirectly through the tuning of quark masses to the physical point. The QED and strong-isospin breaking corrections to the pure QCD result must be calculated with quark masses tuned to the same experimental meson masses. In this way the calculated corrections take the unphysical world of the pure QCD calculation to the physical result. 

The BMW collaboration calculated both of these effects to first order in the small parameters $\alpha_{\mathrm{QED}}$ and $m_d-m_u$ and included them in their determination of the full HVP~\cite{Borsanyi:2020mff}. They found that QED effects lower by $-0.20$\%, and strong-isospin breaking effects increase by +0.27\% the full HVP, largely cancelling out. Results from RBC/UKQCD~\cite{RBC:2018dos} are consistent with this picture, albeit with larger uncertainties. Further lattice-QCD results with an uncertainty comparable to that of BMW are needed. Here we make use of the BMW results to estimate an uncertainty from omitting these effects for our windowed HVP values. 

First, however, we discuss how QED and strong-isospin breaking effects arise. This will allow us to see how well the BMW results agree with expectations and make a physically motivated estimate of the error from omitting these effects for our windows in Table~\ref{tab:results}. 
Our discussion relies on the impact of QED and strong-isospin breaking on the determination of the lattice spacing being very small. This should be true provided that quantities insensitive to these effects are used to fix $a$. BMW demonstrate this to 0.1\% for the $\Omega$ baryon mass~\cite{Borsanyi:2020mff}, but it should also be true for the lattice spacing used here, which is determined from $f_{\pi}$~\cite{Dowdall:2013rya} and includes an estimate of the impact of QED in its uncertainty. 

\subsubsection{QED}
The largest QED effect on the HVP is expected to come from the electric charge of the valence quarks. This is borne out in the BMW calculation~\cite{Borsanyi:2020mff} which finds the impact of QED for the sea quarks and mixed sea/valence effects to be negligible (consistent with zero within their uncertainties). We will therefore focus our discussion on valence (quenched) QED effects. The calculation in Ref.~\cite{Hatton:2020qhk} of the impact of QED on the $c$ quark HVP contribution allows a detailed analysis because the correlation functions can be calculated so precisely. Using a stochastic approach to quenched QED~\cite{Duncan:1996xy} enables a direct comparison of correlation functions with and without the inclusion of QED on the same gluon field configurations. At fixed valence quark mass in lattice units, the effect of QED is to increase the value of the connected correlation function at small values of $t$ and to reduce it at large values of $t$. The latter effect is a result of the QED self-energy contribution exceeding the Coulomb attraction between the quark and antiquark~\cite{Hatton:2020lnm}, thereby pushing up the ground-state meson mass. This is largely offset by the retuning of the $c$-quark mass in the presence of QED so that the $J/\psi$ meson mass (used for tuning) has the same value. QED also increases the $J/\psi$ decay constant, which determines the ground-state amplitude in the $cc$ correlation function. Because the correlation function falls very rapidly with $t$, the contributions to $a_{\mu}$ peak at small $t$ values. The impact of QED on the $c$ quark contribution to the HVP is therefore small and positive at +0.2\%~\cite{Hatton:2020qhk}. This is numerically irrelevant to the total HVP because the $c$-quark contribution is already small. For the $s$ quark we also expect QED effects to be irrelevant because, although the $s$-quark contribution is larger, its electric charge is smaller. Lattice calculations have found a very small $s$-quark QED effect of around -0.03\%~\cite{Giusti:2017jof,RBC:2018dos}. The effect is negative in this case, because the quark mass is tuned from corresponding pseudoscalar mesons and the QED contribution to the hyperfine splitting has the same sign as the QCD contribution for electrically neutral mesons~\cite{Hatton:2020lnm}, increasing the vector meson mass and reducing the contribution to $a_{\mu}$  from the correlation function at large time. 

Compared with the $c$-quark and $s$-quark cases, the dominant light quark HVP contributions are more sensitive to the effect of QED on their correlation functions and have support from larger $t$ values. One contribution to the $ll$ HVP that we can estimate from chiral perturbation theory is that from $2\pi$ states. Because of spin-statistics only $\pi^+\pi^-$ pairs appear. In the absence of QED the $\pi^+$ has the same mass as the $\pi^0$, but when QED is switched on the $\pi^+$ mass shifts upwards from 135.0 MeV to 139.6 MeV. The effect of this on the connected $ll$ contribution was estimated at $-4.3\times 10^{-10}$ in Ref.~\cite{FermilabLattice:2019ugu}. Because the $\pi\pi$ contribution appears only in the isospin-1 channel, the ratio of connected to disconnected contributions is $-$1/10 from the appropriate electric charge factors~\cite{Chakraborty:2015ugp, DellaMorte:2010aq}. This then gives a shift of $+0.43\times 10^{-10}$ to the disconnected contribution. 

The effect of shifting the $\pi^0$ mass to that of $\pi^+$ on the connected $ll$ contribution is largely cancelled by the impact of the simplest hadronic channel in the vector correlator, $\pi^0\gamma$ that becomes available when QED effects are included. From $R_{e^+e^-}$ analyses this contribution is estimated as $+4.5(1)\times 10^{-10}$~\cite{Keshavarzi:2019abf}. 

For the $\rho$ contribution to the $ll$ connected correlation function, it is the impact of QED on the $\rho^0$ that matters. This is arguably very small given the small difference in mass between $\rho^0$ and $\rho^+$ (0.1(1)\%~\cite{pdg}), when the product of their valence quark and antiquark electric charges has opposite sign. On the basis of the $s$-quark results we might expect a small decrease in the connected contribution from QED because of the small increase in the $\rho^0$ mass. Estimating this at -0.15\% (5 times that of $s$ to allow for $u$ and $d$ electric charge factors) of the total $\rho$ contribution, estimated at $425\times 10^{-10}$ from the chiral model of Ref.~\cite{Chakraborty:2016mwy}, gives $-0.6\times 10^{-10}$. 

QED contributions to the $ll$ disconnected correlation function come from dressing the two quark loops with photons but also including photons, along with gluons, connecting the two quark loops. These will give further contributions, beyond shifting the $\pi^0$ mass to that of $\pi^+$, which could be estimated at roughly $e^2/g^2$ times the disconnected contribution itself, i.e. roughly $0.1\times 10^{-10}$ combined with factors of electric charges. The sign of these pieces is not clear, however. 

 Adding together the numbers above, we arrive at rough estimates for QED effects of $-0.4\times 10^{-10}$ for the connected contribution and $\mathcal{O}(0.1\times 10^{-10})$ for the disconnected contribution. Given the approximations involved, our estimates are in reasonable agreement with the BMW results of $-0.87(60)\times10^{-10}$ and $-0.58(18)\times10^{-10}$ (including valence, sea and mixed QED effects) respectively~\cite{Borsanyi:2020mff}. They certainly argue that QED effects on the HVP amount to less than 1\% and are likely to be negative for the $ll$ connected correlation function. 
 
We note that currently there is mild tension between the BMW QED valence quark contribution~\cite{Borsanyi:2020mff} and those from ETMC~\cite{Giusti:2017jof} and RBC/UKQCD~\cite{RBC:2018dos}, which both have larger uncertainties. The sum of QED valence and sea contributions is consistent between BMW and RBC/UKQCD.  Further lattice calculations of QED effects with small enough uncertainties are needed to reach a consensus on these.

 \begin{figure}
    \includegraphics[scale=0.5]{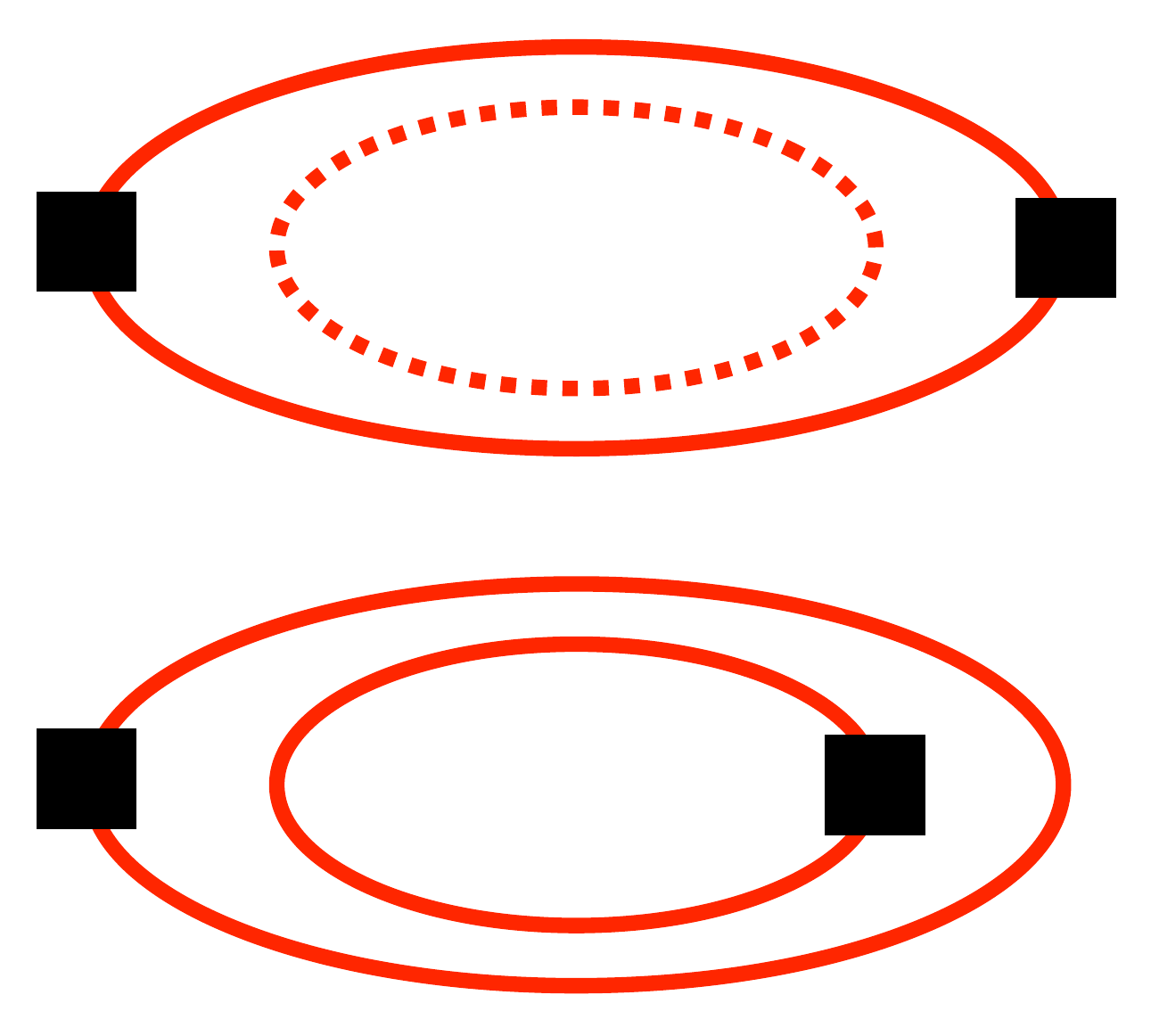}
    \caption{Top: the $\pi\pi$ contribution to the HVP for the connected correlation function in a theory with 1 quark flavour. The solid red line shows a valence quark loop and the dashed red line a sea quark loop. Gluon lines are not shown. Black squares show  insertion points for the electromagnetic current. Bottom: the $\pi\pi$ quark-line disconnected contribution. }
    \label{fig:conn-disc}
\end{figure}

\subsubsection{Strong-isospin breaking (SIB)}
Strong-isospin breaking effects, unlike QED corrections, must make a positive contribution to the HVP because the quark with largest electric charge, $u$, becomes lighter. The effects were recently calculated in chiral perturbation theory in Ref.~\cite{James:2021sor} to be $+3.32(89)\times 10^{-10}$, which should be compared to BMW's value of $+1.93(1.20)\times10^{-10}$~\cite{Borsanyi:2020mff}. Again, this indicates that the sign and size of these corrections found in the lattice calculation are reasonable. The individual quark-line connected and disconnected contributions from strong-isospin breaking have opposite signs (unlike the QED case above) and tend to cancel. 
This is expected because the sizeable contributions from $\pi\pi$ states to the strong-isospin breaking correction have equal magnitude and opposite sign (to next-to-leading-order in chiral perturbation theory) for connected and disconnected pieces~\cite{Lehner:2020crt}. This is easy to see from simple arguments as we show below. 

In a theory with a single quark flavour, the connected and disconnected contributions to the HVP from $\pi\pi$ states must cancel because spin-statistics allows no $\pi\pi$ contribution to the total. Figure~\ref{fig:conn-disc} schematically shows these two contributions in a way that helps to demonstrate this. The same cancellation occurs with two flavours, but we now have to count $\pi$ species and keep track of electric charge factors. Thus we can write 
\begin{eqnarray}
a_{\mu}^{\mathrm{conn}}(\pi\pi) &=& \frac{5}{9}a_{\mu}(\pi^+\pi^-)+\frac{4}{9}a_{\mu}(\pi^0_u\pi^0_u) + \frac{1}{9}a_{\mu}(\pi^0_d\pi^0_d) \\
a_{\mu}^{\mathrm{disc}}(\pi\pi) &=& \frac{4}{9}a_{\mu}(\pi^+\pi^-) - \frac{4}{9}a_{\mu}(\pi^0_u\pi^0_u) - \frac{1}{9}a_{\mu}(\pi^0_d\pi^0_d) \, , \nonumber
\end{eqnarray}
where $\pi^0_u$ is a $u \overline{u}$ pion and $\pi^0_d$ a $d\overline{d}$ pion. The terms involving purely $u$ or purely $d$ quarks have opposite sign as in Fig.~\ref{fig:conn-disc}; the terms that combine $u$ and $d$ quarks have the same sign and differ by electric charge factors where they connect to the external photons. In the isospin-symmetric limit $a_\mu^{\mathrm{conn}}=10a_\mu(\pi \pi)/9$ and $a_{\mu}^{\mathrm{disc}}=-a_\mu(\pi\pi)/9$~\cite{Chakraborty:2015ugp}. 

The isospin-breaking correction is then $\Delta_{\mathrm{SIB}}=a_{\mu}(ud)-a_{\mu}(ll)$ and we see that 
\begin{eqnarray}
\label{eq:deltasib}
\Delta^{\mathrm{conn}}_{\mathrm{SIB}}(\pi\pi) 
&=&-\frac{5}{9}a_{\mu}(\pi^+\pi^-)+\frac{4}{9}a_{\mu}(\pi^0_u\pi^0_u) + \frac{1}{9}a_{\mu}(\pi^0_d\pi^0_d)  \nonumber \\
&=& -\Delta^{\mathrm{disc}}_{\mathrm{SIB}}(\pi\pi) \, .
\end{eqnarray}
Each of $\Delta^{\mathrm{conn}}_{\mathrm{SIB}}(\pi\pi)$ and $\Delta^{\mathrm{disc}}_{\mathrm{SIB}}(\pi\pi)$ vanish, by definition, in the isospin-symmetric limit when all the $\pi$ masses are the same. Away from this limit $\pi^0_u$ and $\pi^0_d$ will have lighter and heavier masses respectively than $\pi^+$ (whose mass is that of $\pi^0$ when $m_u\ne m_d$ but QED is switched off), but then $\Delta^{\mathrm{conn}}_{\mathrm{SIB}}(\pi\pi)$ and $\Delta^{\mathrm{disc}}_{\mathrm{SIB}}(\pi\pi)$ will cancel, as shown in Eq.~\eqref{eq:deltasib}. This means that the total SIB contribution is dominated by $\rho$-$\omega$ mixing~\cite{James:2021sor}.

BMW~\cite{Borsanyi:2020mff} find separate connected and disconnected SIB contributions to the HVP of $6.60(82)\times 10^{-10}$ and $-4.67(88)\times 10^{-10}$. These results are consistent both with the value of $\Delta^{\mathrm{conn}}_{\mathrm{SIB}}=6.9(3.5)\times10^{-10}$ found for the $\pi\pi$ contribution in Ref.~\cite{Lehner:2020crt} and with the total SIB correction found in Ref.~\cite{James:2021sor}, as well as the earlier lattice-QCD results for the connected SIB contribution in Refs.~\cite{FermilabLattice:2017wgj,RBC:2018dos}.

\subsubsection{QED$+$SIB: net isospin-breaking correction}

We conclude from our analyses that the QED and strong-isospin breaking corrections to the total HVP are both small, of similar size, and with opposite sign (for at least the connected contribution). The almost complete cancellation seen in the BMW results~\cite{Borsanyi:2020mff} might not have been anticipated but is also not surprising. BMW find a total QED correction of $-1.45(62)\times 10^{-10}$, which amounts to $-$0.2(1)\% of the contribution from $ll$ correlators. For strong-isospin breaking the correction is $1.93(1.20)\times 10^{-10}$, or +0.3(2)\%. 

For the strong-isospin breaking correction to our windowed results we expect a very similar relative correction. This is because, from the arguments above, the $\pi\pi$ contributions in connected and disconnected strong-isospin breaking corrections will cancel for any windows, leaving a correction dominated by $\rho-\omega$ mixing. Since the $\rho$ and $\omega$ masses control the time dependence of the correlation functions in our windows, we expect the correction to scale in proportion to the contribution for that window, giving the same percentage correction as above. Indeed, this is what was found by BMW for the smaller window that they used in Ref.~\cite{Borsanyi:2020mff}. 

For the QED correction to our windowed results, we expect somewhat different behaviour. These corrections, as we have discussed above, are dominated by the cancelling $2\pi$ and $\pi^0\gamma$ contributions, which will be suppressed strongly by our windows when large $t$ values are cut out. We therefore expect the QED correction for the windows to be more sensitive to effects on the $\rho$ correlator. As discussed for the $c$ and $s$ cases, the QED effects change from positive to negative as $t$ increases. We therefore expect the (negative) QED effects to become relatively smaller in size for the windowed results. This is again what was found by BMW; for the window they used (extending to $t_1=1.0$\,fm) the QED correction was cut to $-0.04(3)$\%. For our windows with larger $t_1$ we do not expect to see as much of a reduction. RBC/UKQCD give results~\cite{RBC:2018dos} for a variety of window sizes, consistent with this picture. 

BMW~\cite{Borsanyi:2020mff} find the combined QED plus strong-isospin-breaking correction for the window they used to be +0.18(3)\%. The Mainz/CLS~\cite{Ce:2022kxy} QED plus strong-isospin-breaking correction of 0.3(1)\% calculated for the same window for the quark-line connected contribution agrees well with the corresponding result from BMW. 

The arguments above lead to a range of plausible QED plus strong-isospin breaking corrections of between +0.1(2)\% and +0.2(2)\% for our windows, depending on the window size. 
We do not make a correction to our values, but instead add an uncertainty of 0.2\% for these corrections to each of the windowed lattice results. This is given as the second uncertainty for the total $a_{\mu}^w(\mathrm{latt})$ column of Table~\ref{tab:results}.

\subsection{The total lattice-QCD result in each window}
\label{sec:latt-tot}
The different lattice contributions discussed in Secs.~\ref{sec:connll},~\ref{sec:connsc} and~\ref{sec:discwindow} must be added together to arrive at a total for the HVP for each window. The sum is given in column~7 of Table~\ref{tab:results}. The first uncertainty is that coming from all of the contributions tabulated and is dominated by that from the connected $ll$ contribution. The second uncertainty comes from the missing QED and strong-isospin breaking corrections discussed in Sec.~\ref{sec:QEDSIB}. The advantage of the windowing is clear from these numbers; the relative uncertainty on $a_{\mu}^w(\mathrm{latt})$ is reduced from the 1.3\% for the full $a_{\mu}$ found in Ref.~\cite{FermilabLattice:2019ugu} (where variance reduction techniques were used for the $ll$ connected contribution that are not needed here) to 0.7\% for the time-window with $t_1=1.5$\,fm.

\section{The HVP from $R_{e^+e^-}$ with a time-window} 
\label{sec:R}

Experimental data on the cross-section for $e^+e^-$~annihilation into hadrons and $R_{e^+e^-}$, the cross-section ratio between hadrons and muons in the final state, are available from several different experiments and for many different hadronic channels that open up once the centre-of-mass energy $\sqrt{s}>m_{\pi^0}$~\cite{Bai:1999pk,Akhmetshin:2000ca,Akhmetshin:2000wv,Achasov:2000am,Bai:2001ct,Achasov:2002ud,Akhmetshin:2003zn,Aubert:2004kj,Aubert:2005eg,Aubert:2005cb,Aubert:2006jq,Aulchenko:2006na,Achasov:2006vp,Akhmetshin:2006wh,Akhmetshin:2006bx,Akhmetshin:2006sc,Aubert:2007ur,Aubert:2007ef,Aubert:2007uf,Aubert:2007ym,Akhmetshin:2008gz,Ambrosino:2008aa,Ablikim:2009ad,Aubert:2009ad,Ambrosino:2010bv,Lees:2011zi,Lees:2012cr,Lees:2012cj,Babusci:2012rp,Akhmetshin:2013xc,Lees:2013ebn,Lees:2013uta,Lees:2014xsh,Achasov:2014ncd,Aulchenko:2014vkn,Akhmetshin:2015ifg,Ablikim:2015orh,Shemyakin:2015cba,Anashin:2015woa,Achasov:2016bfr,Achasov:2016lbc,TheBaBar:2017aph,CMD-3:2017tgb,TheBaBar:2017vzo,Kozyrev:2017agm,Anastasi:2017eio,Achasov:2017vaq,Xiao:2017dqv,TheBaBar:2018vvb,Anashin:2018vdo,Achasov:2018ujw,Lees:2018dnv,CMD-3:2019ufp}.
To determine the HVP contribution to $a_{\mu}$, the results must be collated into values for $R_{e^+e^-}$ as a function of $\sqrt{s}$, summing over exclusive channels, reconciling different experimental results, allowing for correlations and applying QED radiative corrections. This is a challenging task; see Ref.~\cite{Aoyama:2020ynm} for a discussion of how this is done. A few different groups carry out this work~\cite{Benayoun:2019zwh, Davier:2019can, Keshavarzi:2019abf}, using a variety of approaches to combine the experimental data and estimate errors. Overall the agreement among the different groups is good although there is some variation between results and uncertainties for particular channels. Here we will use, as an example, the results for $R_{e^+e^-}$ from KNT19~\cite{Keshavarzi:2019abf}.\footnote{These were kindly supplied to us by Alex Keshavarzi. The data set extends to just above 11\,GeV. We use $\mathcal{O}(\alpha_s^2)$ perturbation theory for higher energies; higher orders are negligible.} 

The determination of the HVP from $R_{e^+e^-}$ generally proceeds by integration over $\sqrt{s}$ using a kernel function~\cite{Aoyama:2020ynm}, 
\begin{equation}
\label{eq:intR}
a_{\mu}^{\mathrm{HVP}}=\left(\frac{\alpha^2}{3\pi^2}\right) \int_{m_{\pi}^2}^{\infty} ds \frac{R_{e^+e^-}}{s} K_R(s) \, .
\end{equation}
For our purposes here, however, it is more convenient to transform the results for $R_{e^+e^-}$ into a `lattice correlation function' as a function of Euclidean time, $G_R(t)$, using~\cite{Bernecker:2011gh}:
\begin{equation}
    \label{eq:GR}
    G_R(t) \equiv \frac{1}{12\pi^2} \int_0^\infty\!dE\,E^2\, R_{e^+e^-}\,\mathrm{e}^{-E|t|} \,,
\end{equation}
where $E=\sqrt{s}$ is the centre of mass energy.
We evaluate the integral by fitting the integrand to a monotonic (Steffen) spline and integrating the spline function exactly~\cite{Steffen:1990}. This choice gives slightly more accurate results than the Trapezoidal Rule, but the difference is negligible.
We evaluate $G_R(t)$ for a discrete set of $t$, corresponding to a very fine lattice spacing. We can then manipulate it in the same way as the lattice data and obtain the partial HVP contribution for each window.\footnote{We have checked that we obtain the same result and uncertainty as KNT19 for the full HVP contribution when doing these manipulations.} 
The partial HVP results for each $t_1$ value for the windows of Sec.~\ref{sec:lattice} are given in Table~\ref{tab:results}. 

\begin{figure}
    \parbox[center]{0.92\linewidth}{
    \begin{flushright}   
    \includegraphics[scale=0.9]{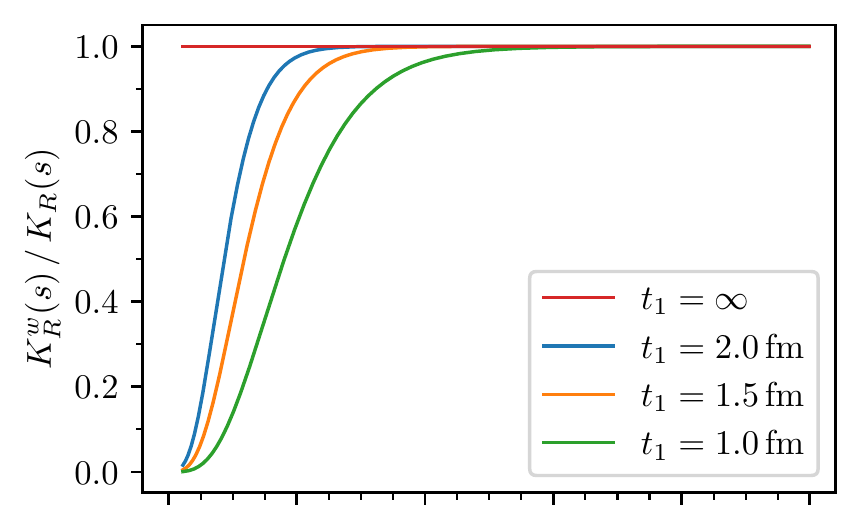} \\
    \vspace{-2ex}
    \includegraphics[scale=0.9]{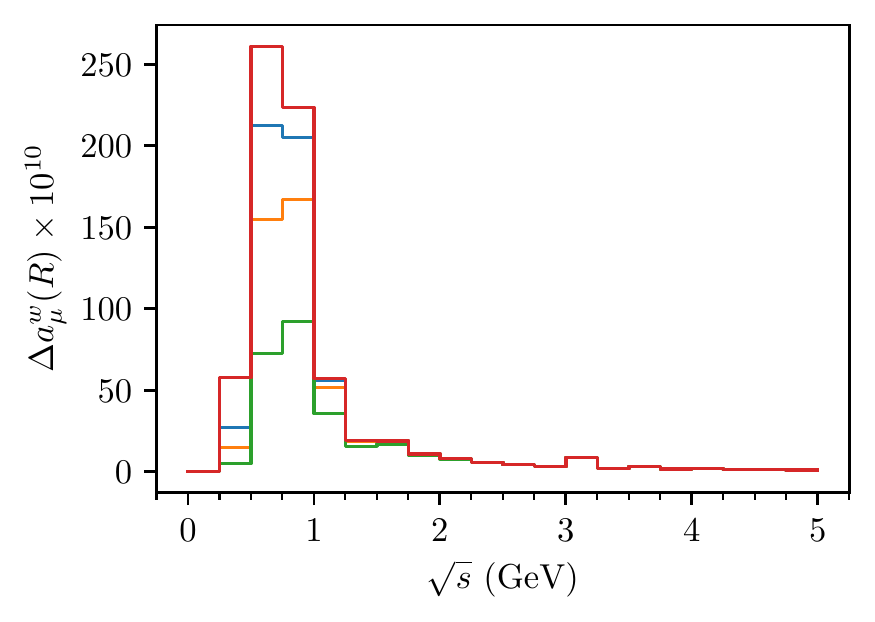}
    \end{flushright}
    }\vspace{-3ex}
\caption{Top: Ratio of effective kernels, $K_R^w(s)$, from the integral to obtain
    $a_\mu^w$ (Eq.~\eqref{eq:intRw}) to $K_R(s)$ for $a_{\mu}$ as a function of $\sqrt{s}$ (Eq~\eqref{eq:intR}). Results are for our one-sided time-windows where 
    (upper curve to lower curve) $t_1=\infty$ (red), $t_1=2.0$ (blue), 1.5 (orange), and 1.0 (green)\,fm and $\Delta t =$0.15\,fm. 
    Bottom: Integrand $\Delta a_\mu^w$ of Eq.~\ref{eq:intwin} from $R_{e^+e^-}$ in bins of width 
    0.25\,GeV in $\sqrt{s}$. Results are shown for
   the one-sided time-windows of the top pane with corresponding colours.}
    \label{fig:KRw}
\end{figure}

Although the impact of the time-window is easily visualised in $t$-space (see Fig.~\ref{fig:win-t}), it is less clear what its impact is when translated to $\sqrt{s}$-space. The quantity we need is the effective kernel~$K_R^w(s)$ corresponding to a time window with kernel $K_G^w(t)$, where
\begin{equation}
    \label{eq:intRw}
    a_{\mu}^{w}=\left(\frac{\alpha^2}{3\pi^2}\right) \int_{m_{\pi}^2}^{\infty} ds \frac{R_{e^+e^-}}{s} K^w_R(s) \,. 
\end{equation}
Here we obtain this by differentiating (numerically, using automatic differentiation) the value of $a_\mu^w$, obtained from Eqs.~(\ref{eq:GR}) and~(\ref{eq:intwin}), with respect to individual input values for $R_{e^+e^-}$ at different energies~$\sqrt{s}$ (but see also Eq.~(184) in Ref.~\cite{Borsanyi:2020mff}). The upper pane of Fig.~\ref{fig:KRw} shows the ratio of kernel $K^w_R$ to $K_R$ for 3 different values of $t_1$, demonstrating the effect of the time-window in $\sqrt{s}$-space. The impact of introducing, and then reducing, $t_1$ is to successively cut out more of the  low $\sqrt{s}$ region. Contributions from~$\sqrt{s}$ below 0.5\,GeV are significantly affected for the largest time-window, $t_1 =$2\,fm, while the smallest time-window suppresses contributions from $\sqrt{s}$ below 1.25\,GeV. Contributions from values of $\sqrt{s}$ above 1.25\,GeV are scarcely affected for any of the time-windows. 

The lower pane of Fig.~\ref{fig:KRw} shows binned contributions to the HVP from $R_{e^+e^-}$ versus $\sqrt{s}$, with and without the application of time-windows. Consistent with the kernel functions plotted in the upper pane, the smaller time windows substantially suppress contributions from $\sqrt{s}< $1\,GeV, while barely affecting contributions from $\sqrt{s} > 1.5\,$GeV.

The dominant contributions to the HVP from $R_{e^+e^-}$ in the $\sqrt{s} < 1$\,GeV region are from the exclusive channels $\pi^0\gamma$,  $\eta\gamma$, $\pi^+\pi^-$, $\pi^+\pi^-\pi^0$ and $4\pi$~\cite{Keshavarzi:2018mgv}. There is also a spike in $R_{e^+e^-}$ around 1\,GeV from the $\phi$ and $KK$ channels. However, $\sqrt{s} <$ 1\,GeV is below the region where channels with many particles kick in. Thus imposing our one-sided time-window on the $R_{e^+e^-}$ data affects the contribution of only a few number of channels. 

\section{Results and comparison}
\label{sec:results}

We can now compare the results from lattice QCD and from $R_{e^+e^-}$ data for the HVP for our one-sided time-windows, using the results shown in columns 7 and 8 in Table~\ref{tab:results}.  The partial contribution to the HVP from each time-window is a physical quantity.  This means that the windowed lattice and $R_{e^+e^-}$ results should agree for all values of $t_1$ and not just as $t_1 \rightarrow \infty$ (for the full HVP)~\cite{Bernecker:2011gh, Lehner:2017kuc}. A significant difference between lattice and $R_{e^+e^-}$ results for any value of $t_1$ is a sufficient condition to raise issues for the SM determination of the HVP contribution to $a_{\mu}$. 

\begin{figure}
    \includegraphics[scale=0.9]{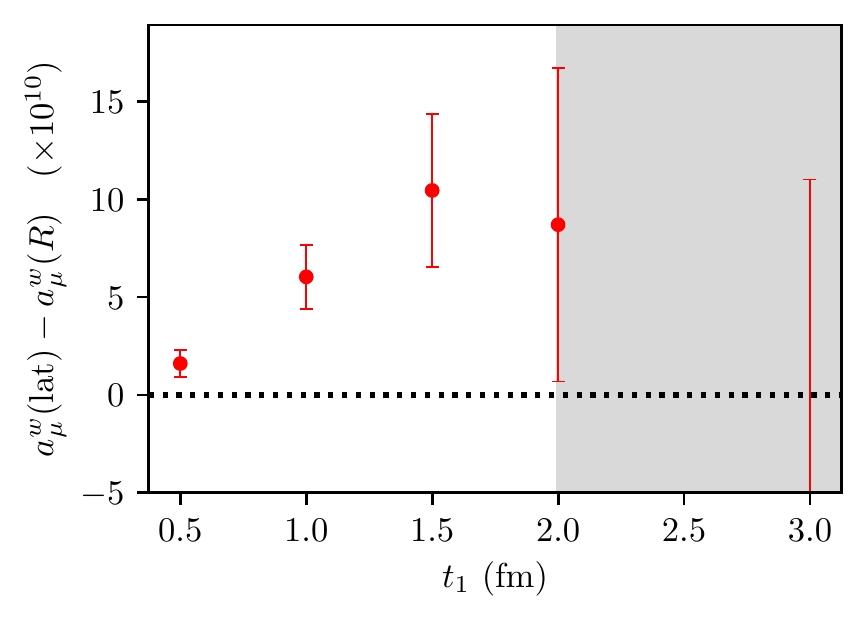}
    \caption{Difference between determinations of 
    $a_\mu^w$ from the lattice and from $R_{e^+e^-}$
    with one-sided windows for different values 
    of $t_1$. Results for $t_1=1.0$ and $1.5$\,fm 
    differ from zero by 3.7$\sigma$ and 2.7$\sigma$, respectively.
    We have insufficient statistics to give reliable results for $t_1>2$\,fm (grey shading).}
    \label{fig:diff}
\end{figure}

\begin{figure}
    \includegraphics[scale=0.9]{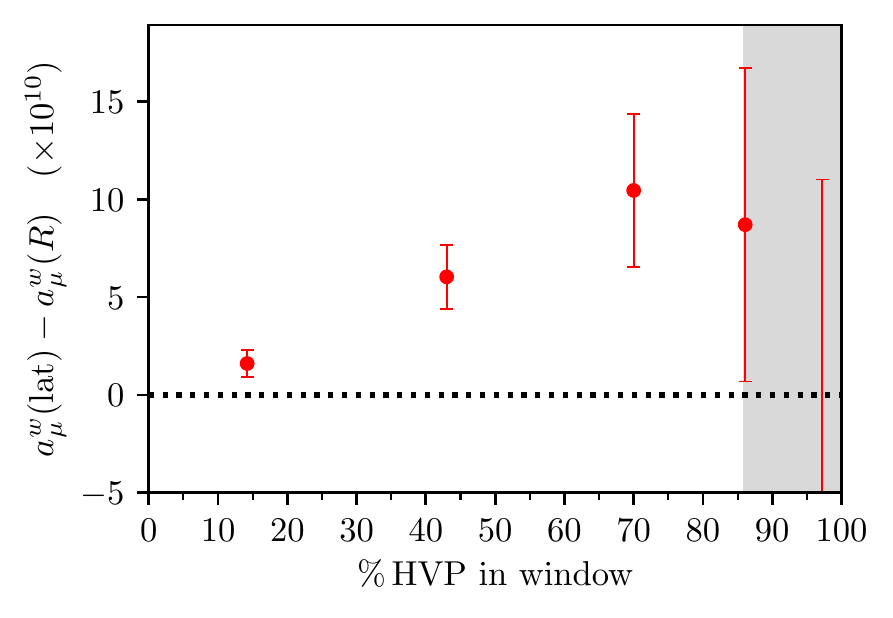}
    \caption{Difference between determinations of 
    $a_\mu^w$ from the lattice and from $R_{e^+e^-}$
    with one-sided windows for (from left to right) 
    $t_1=0.5$, 1, 1.5, 2 and~3\,fm.  The 
    differences are plotted versus the fraction of the total
    HVP included in the window. We have insufficient statistics to give reliable results for $t_1>2$\,fm (grey shading).}
    \label{fig:diffvsfrac}
\end{figure}

\begin{figure}
    \includegraphics[scale=0.9]{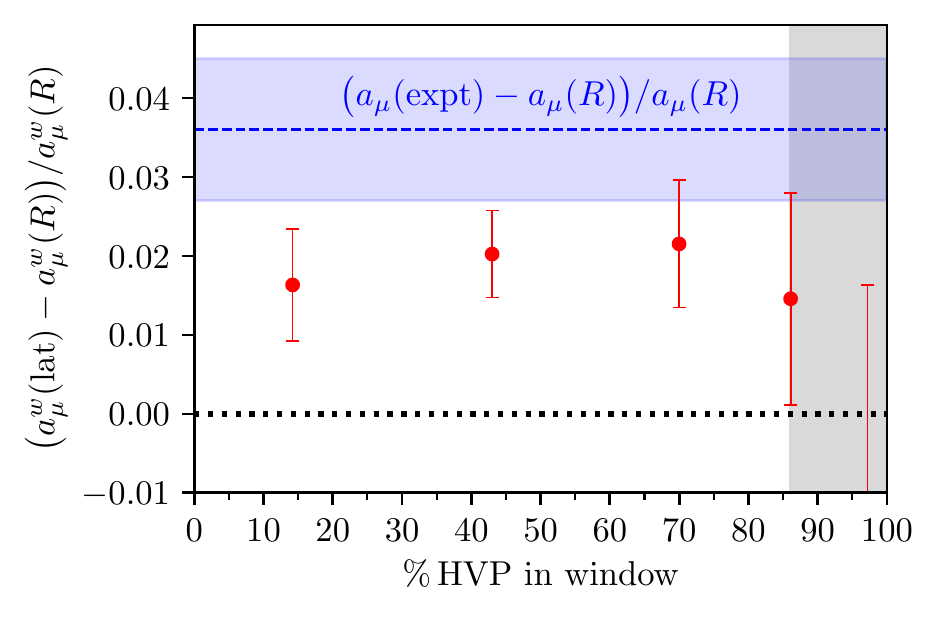}
    \caption{Fractional difference between determinations of 
    $a_\mu^w$ from the lattice and from $R_{e^+e^-}$
    with one-sided windows for different values 
    of $t_1$. The differences are plotted versus the fraction of the total
    HVP included in the window. We have insufficient statistics to give reliable results for $t_1>2$\,fm (grey shading). 
    For comparison, the current difference between the 
    experimental average for $a_{\mu}$ and the SM~$a_\mu$ using the data-driven HVP contribution divided by the SM $a_{\mu}$ is 0.036(9) (blue band).
}
    \label{fig:reldiff}
\end{figure}

The difference $a_{\mu}^w(\mathrm{latt})-a_{\mu}^w(R)$ is  reported in column 9 of Table~\ref{tab:results} and plotted as a function of $t_1$ in Fig.~\ref{fig:diff}. The lattice results are higher than those from $R_{e^+e^-}$ for $t_1 \le 2$\,fm, so that $a_{\mu}^w(\mathrm{latt})-a_{\mu}^w(R)$ is positive. The difference rises as a function of $t_1$ to values around $10 \times 10^{-10}$ by $t_1 = 1.5$\,fm and points to tension between these two results. Beyond this point the lattice results become noisy and the uncertainty of the difference becomes large with low statistical significance. Improving the results at $t_1=1.5$ and~$2$\,fm is crucial for establishing whether the growth with $t_1$ is correct and whether the difference grows further beyond~1.5\,fm.

Figure~\ref{fig:diffvsfrac} is the same as Fig.~\ref{fig:diff} but plotted versus the fraction of the total HVP included in the window (\%\,HVP in Table~\ref{tab:results}) instead of~$t_1$. The results with $t_1\ge1.5$\,fm include much larger fractions of the HVP than the smaller values.

\begin{table}
    \caption{\label{tab:errbudget} Percent errors in $a_\mu^w(\mathrm{latt})-a_\mu^w(R)$ 
    (last column of Table~\ref{tab:results}) from different sources for one-sided windows with $t_1=1$,~1.5 and~2\,fm. The sources are in order of decreasing importance for the $t_1=1.5$\,fm case as one goes down the list. Error sources include: Monte Carlo noise in the correlators from the lattice simulations, uncertainties in the lattice spacing and the renormalization constant for the vector currents, experimental uncertainty in $R_{e^+e^-}$,  the uncertainty in the disconnected contribution ($(ll+ss)_\mathrm{disc}$ in Table~\ref{tab:results}), an uncertainty to account for contributions from QED and strong isospin breaking,  uncertainties in corrections that remove effects from mistuned pion masses and the lattice's finite volume, uncertainty in the 
    extrapolation to zero lattice spacing~$a$,   uncertainties in the $s$, $c$, and $b$ (connected) vacuum polarizations (Table~\ref{tab:results}), and tuning uncertainties in the sea quark masses.
    }
    \begin{ruledtabular}\begin{tabular}{lrrr}
        $t_1$  (fm) & 1.0\phantom{\%} & 1.5\phantom{\%} & 2.0\phantom{\%} \\
        \hline
        Statistics                     &  8.4\phantom{\%} & 22.0\phantom{\%} & 56.3\phantom{\%} \\
        Lattice spacing, $Z_V$         &  8.8\phantom{\%} & 20.6\phantom{\%} & 37.8\phantom{\%} \\
        $R_{e^+e^-}$                   & 21.1\phantom{\%} & 17.1\phantom{\%} & 19.8\phantom{\%} \\
        Disconnected vac. pol.         &  3.7\phantom{\%} &  9.9\phantom{\%} & 20.6\phantom{\%} \\
        QED, strong isospin breaking   & 10.4\phantom{\%} &  9.8\phantom{\%} & 11.6\phantom{\%} \\
        Stagg. pions/finite vol.       &  5.2\phantom{\%} &  8.6\phantom{\%} & 12.6\phantom{\%} \\
        $a\to0$ extrapolation          &  1.2\phantom{\%} &  3.0\phantom{\%} & 15.8\phantom{\%} \\
        $s$-quark vac. pol.            &  2.7\phantom{\%} &  2.8\phantom{\%} &  3.2\phantom{\%} \\
        Sea masses                     &  0.8\phantom{\%} &  1.3\phantom{\%} &  1.6\phantom{\%} \\
        $c$,$b$-quark vac. pol.        &  0.9\phantom{\%} &  0.5\phantom{\%} &  0.5\phantom{\%} \\
        \hline
        Total                          & 27.5\% & 38.5\% & 77.3\% \\
    \end{tabular}\end{ruledtabular}
\end{table}

 Table~\ref{tab:errbudget} provides error budgets for the difference $a_\mu^w(\mathrm{latt})-a_\mu^w(R)$ with 
different values of~$t_1$ (see Ref.~\cite{FermilabLattice:2019ugu} for more details on the underlying  analysis). 
As expected, the importance of statistical errors decreases significantly as $t_1$~decreases. At the same time the relative uncertainty due to~$R_{e^+e^-}$ increases. Discretization errors also decrease with decreasing~$t_1$, as might be expected from the behaviour evident in Fig.~\ref{fig:t1vsa}.

For one-sided windows with $t_1 = 1$ and~$1.5$\,fm the differences between lattice QCD and $R_{e^+e^-}$ amount to tensions of~3.7 and~2.7 standard deviations, respectively. These are marginally statistically significant, but the error budget suggests that increasing lattice statistics by a factor of 5--10 would shrink the total uncertainty in $a_{\mu}^w(\mathrm{latt})$  for $t_1=1.5$ and 2\,fm substantially, particularly if a value for~$w_0$ with smaller uncertainty is obtained. The errors would then be comparable to the errors in~$a_\mu^w(R)$. The errors at $t_1=1$\,fm will be harder to improve because they are dominated by uncertainties in~$R_{e^+e^-}$.

We emphasize here that we are giving an example of the analysis possible; a more complete analysis would be needed to clarify the significance of the results for $a_{\mu}^w(\mathrm{latt})-a_{\mu}^w(R)$. We use the $R_{e^+e^-}$ results from KNT19~\cite{Keshavarzi:2019abf}. In Ref.~\cite{Aoyama:2020ynm} a more conservative uncertainty estimate is quoted, which allows for different possible treatments of the underlying cross-section data (see also Ref.~\cite{Davier:2019can}). Taking that approach here (see Ref.~\cite{Colangelo:2022vok}) would increase the uncertainty on $a_{\mu}^w(R)$.  This would not have a large effect on $a_{\mu}^w(\mathrm{latt})-a_{\mu}^w(R)$ because its uncertainty is dominated by that from lattice QCD except at very small $t_1$ (see Table~\ref{tab:results}). On the lattice side our analysis is incomplete because we are missing a full set of correlators that would allow us to determine QED and strong-isospin breaking effects. For these we take estimates based on results from the BMW collaboration~\cite{Borsanyi:2020mff}. Even for the full HVP they find these effects each to be small, $\mathcal{O}(2\times 10^{-10})$, and tending to cancel. We discuss why this happens in Section~\ref{sec:QEDSIB} and reason that cancellations should persist under the application of time-windows. We then take an uncertainty of 0.2\% for these corrections, which is double the relative effect seen in the full $a_{\mu}$. We also use BMW results~\cite{Borsanyi:2020mff} to normalise the quark-line disconnected contribution, because the results of Ref.~\cite{FermilabLattice:2021hzx} are blinded. The BMW results agree well with a recent data-driven determination~\cite{Boito:2022rkw} (see Sec.~\ref{sec:discwindow}). The disconnected contribution is relatively small, at less than 1\% for windows with $t_1=1.5$\,fm, so the exact numerical value also makes little difference to the significance of our results. 

Figure~\ref{fig:reldiff} plots the size of $a_{\mu}^w(\mathrm{latt})-a_{\mu}^w(R)$ relative to $a_{\mu}^w(R)$ to give a clearer picture of how the tension between the two results changes as we increase the proportion of the HVP included in $a_{\mu}^w$ with increasing $t_1$. 
Our results give a fairly flat curve, at the level of the uncertainties that we have. A growing relative tension would indicate that the tension was being driven by the low $\sqrt{s}$ region.

\begin{figure}
    \includegraphics[scale=0.9]{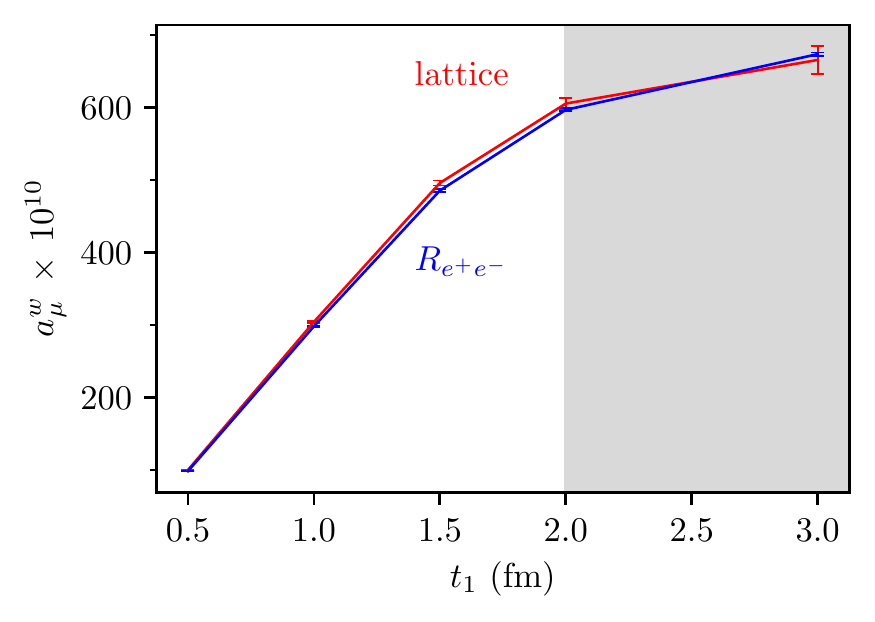}
    \caption{Determinations of 
    $a_\mu^w$ from the lattice (top, red) and from $R_{e^+e^-}$ (bottom, blue)
    with one-sided windows for different values 
    of $t_1$. We have insufficient statistics to give reliable results for $t_1>2$\,fm (grey shading).}
    \label{fig:both}
\end{figure}

Figure~\ref{fig:both} plots the results for $a_{\mu}^w$ from lattice QCD and from $R_{e^+e^-}$ as a function of $t_1$ for direct comparison. The lattice results pull away from the $R_{e^+e^-}$ results upwards as $t_1$ is increased, until $t_1 =$ 2\,fm when the lattice results become too noisy for a useful comparison (and, as before, we have greyed out that region).

\section{Conclusions}
\label{sec:conclusions}

The aim of this work is to investigate the efficacy of one-sided time windows to maximise the significance of any tensions between lattice results and between lattice and data-driven results for the HVP, given current limitations on the statistical and systematic errors on the lattice data at large Euclidean times. Using correlators previously analysed in Refs.~\cite{FermilabLattice:2019ugu,Chakraborty:2014mwa}, supplemented with additional (still preliminary) results for quark-line disconnected contributions~\cite{FermilabLattice:2021hzx} and discussion of missing QED/SIB contributions (see Sec.~\ref{sec:QEDSIB}), we show that a one-sided time window provides a partial result for the HVP contribution to $a_{\mu}$ from lattice QCD with a much smaller uncertainty than that for the full result. This feature agrees with earlier results using a time-window~\cite{Lehner:2017kuc} but, whereas they included only one-third of the full HVP, our time-window  (for $t_1 = $1.5\,fm) corresponds to a large (70\%) fraction of the full result. Using the lattice-QCD correlation functions from the 2019 analysis of Ref.~\cite{FermilabLattice:2019ugu}, we find that the partial HVP result for $t_1 =$1.5\,fm is already determined to better than 1\% uncertainty in this first analysis. 

For this partial result we find a 2.7$\sigma$ tension with the corresponding result from $R_{e^+e^-}$.  See Sec.~\ref{sec:results} for a detailed discussion of the error budget for the difference of windowed HVP values and the limitations of our analysis. We make use of BMW's results for the disconnected and QED+SIB contributions; a completely independent analysis of these effects with comparable precision is very desirable.  Our result is consistent with the effect seen in earlier results with a smaller window by the BMW~\cite{Borsanyi:2020mff}, Mainz/CLS~\cite{Ce:2022kxy} and ETM~\cite{Alexandrou:2022amy} collaborations, but it is also consistent with results where less tension was seen from the RBC/UKQCD collaboration~\cite{RBC:2018dos}; results with higher statistics are needed. A statistically significant difference of HVP values between lattice and data-driven approaches for one-sided windows with $t_1=1.5$\,fm would raise issues for the SM determination of the HVP. 

The urgent question at issue is whether or not the total HVP obtained from lattice QCD agrees with that from data-driven methods. A larger value for the total HVP increases the SM $a_{\mu}$, moving it closer to the average experimental value. The direction of the tension seen with our windowed results (that include 70\% of the HVP) is positive; they are of marginal significance, however. Smaller uncertainties and wider windows will be needed to reach clearer conclusions. Our analysis shows that a result that is statistically significant compared with the current tension between the SM with data-driven HVP and experimental results for $a_{\mu}$ (4\% of the HVP) is within reach of current lattice simulations for $t_1=2$\,fm, which includes 86\% of the HVP.  Obtaining such results for $t_1=2.5$\,fm, which includes 94\% of the HVP, may also be feasible. Results for one-sided time-windows could then rule out or find a difference between lattice QCD and data-driven approaches to the HVP with the uncertainty needed for clarity on $a_{\mu}$. Further analysis also needs more tests of systematic errors in both lattice QCD and data-driven approaches and these are underway. 

The result given here for the data-driven windowed HVP uses the KNT19 analysis~\cite{Keshavarzi:2019abf}. While we expect effects to be small, accounting for different treatments of underlying cross-section data, as in Refs.~\cite{Aoyama:2020ynm,Davier:2019can,Colangelo:2022vok}, would provide a better quantification of the data-driven uncertainty. More experimental data is expected for the data-driven approach with improved uncertainties; this should help address tensions between experimental measurements in important exclusive channels. Tests of existing data in low multiplicity channels against analyticity, unitarity and crossing symmetry constraints provide important constraints~\cite{Colangelo:2018mtw,Hoferichter:2019mqg,Aoyama:2020ynm}. Results from the MuonE experiment~\cite{CarloniCalame:2015obs,Abbiendi:2016xup,Banerjee:2020tdt, Abbiendi:2022oks} should provide useful complementary input on the vacuum polarisation function at space-like $q^2$ values. 

On the lattice-QCD side, we believe that the one-sided windows presented here provide an excellent basis for comparison of lattice HVP results and for lattice to $R_{e^+e^-}$ comparison. We hope that other lattice-QCD collaborations will provide results for these windows, to enable stringent comparisons and so that a robust consensus can be reached. We note that our result for $t_1=1.0$\,fm, where we have a tension of 3.7$\sigma$ with KNT19,  agrees well with the equivalent sum of ETM results~\cite{Alexandrou:2022amy} for the short-distance and intermediate-distance windows.

Meanwhile the Fermilab/HPQCD/MILC  collaborations are improving their HVP results beyond those presented in Ref.~\cite{FermilabLattice:2019ugu} with higher statistics and a blinded analysis, inclusion of QED effects and strong-isospin breaking, and an improved analysis of the large-time behaviour of the $u/d$ connected correlation function~\cite{Lahert:2021xxu}. The aim is a full HVP result with an uncertainty of less than 1\%. In the shorter term, however, we expect to improve uncertainties on the windowed values for the HVP, thereby increasing the largest $t_1$ value that can be used to above 2\,fm. A larger lever arm in $t_1$, along with reduced uncertainties, will provide a clearer map of the region in $\sqrt{s}$ to which any tension seen (if it remains) can be traced.

\subsection*{\bf{Acknowledgements}} 
We are grateful to Alex Keshavarzi for useful discussions on the KNT19 results. This work used the DiRAC Data Analytic system at the University of Cambridge, operated by the University of Cambridge High Performance Computing Service on behalf of the STFC DiRAC HPC Facility (www.dirac.ac.uk). This equipment was funded by BIS National E-infrastructure capital grant (ST/K001590/1), STFC capital grants ST/H008861/1 and ST/H00887X/1, and STFC DiRAC Operations grant ST/K00333X/1. DiRAC is part of the National E-Infrastructure.
We are grateful to the Cambridge HPC support staff for assistance.
We thank the University of Plymouth for providing computing time on the local HPC cluster. 
Computations for this work were also carried out with resources provided by the USQCD Collaboration, the National
Energy Research Scientific Computing Center and the Argonne Leadership Computing Facility, which are funded
by the Office of Science of the U.S.\ Department of Energy.
{This work used the Extreme Science and Engineering Discovery Environment (XSEDE) supercomputer Stampede 2 at the Texas Advanced Computing Center (TACC) through allocation TG-MCA93S002.  The XSEDE program is supported by the National Science Foundation under grant number ACI-1548562.}
{Computations on the Big Red II+ supercomputer were supported in part by Lilly Endowment, Inc., through its support for the Indiana University Pervasive Technology Institute.} The parallel file system employed by Big Red II+ is supported by the National Science Foundation under Grant No.~CNS-0521433.
This work utilized the RMACC Summit supercomputer, which is supported by the National Science Foundation (awards ACI-1532235 and ACI-1532236), the University of Colorado Boulder, and Colorado State University. The Summit supercomputer is a joint effort of the University of Colorado Boulder and Colorado State University.
Some of the computations were done using the Blue Waters sustained-petascale computer, which was supported by the National Science Foundation (awards OCI-0725070 and ACI-1238993) and the state of Illinois. Blue Waters was a joint effort of the University of Illinois at Urbana-Champaign and its National Center for Supercomputing Applications.

Funding for this work came from the
UK Science and Technology Facilities Council (grant ST/T000945/1), the Department of Energy (awards DE-SC0015655, DE-SC0010120 and DE-SC0010005), the
National Science Foundation (grants PHY17-19626 and PHY20-13064) and from their Graduate Research Fellowship (under grant DGE 2040434) and from the Universities Research Association (Visiting Scholarship award 21-S-05). This document was prepared using the resources of the Fermi National Accelerator Laboratory (Fermilab), a U.S. Department of Energy, Office of Science, HEP User Facility. Fermilab is managed by Fermi Research Alliance, LLC(FRA), acting under Contract No.~DE-AC02-07CH11359.

\begin{appendix}

\section{Quark-line disconnected contribution to the lattice-QCD result for the HVP} 
\label{sec:disc} 

The quark-line disconnected contribution to the HVP is relatively small but challenging to calculate. It is most conveniently calculated as the correlation function of two electromagnetic currents, combining all flavours, because it is not flavour-diagonal. Because $c$ and $b$ contributions are negligible, this means combining $u$, $d$ and $s$ quark currents. In the isospin limit being used here we write
\begin{equation}
\label{eq:discj}
J^i_{\mathrm{em}} = \frac{1}{3} (\overline{l}\gamma^i l - \overline{s}\gamma^i s) 
\end{equation}
with the factor $1/3$ coming from the electric charges.
For staggered quarks we must use a `taste-singlet' version of the vector current, which is point-split by one link in the $i$ direction. We normalize the lattice vector current to match that in continuum QCD using a renormalisation factor $Z_V$ determined using the symmetric momentum-subtraction scheme (RI-SMOM) on the lattice~\cite{Hatton:2019gha}. 

The quark-line disconnected correlation function then takes the form 
\begin{equation}
\label{eq:Gdisc}
G^{\mathrm{disc}}(t^{\prime})=\frac{Z_V^2}{T}\sum_t \langle L(t)L(t+t^{\prime})\rangle \, .
\end{equation}
The loop $L(t)$ is constructed from quark propagators as
\begin{equation}
\label{eq:Lt}
L(t) = \frac{1}{3}\mathrm{Tr} \left( \gamma^i \frac{1}{{\not}D+m_l}  -\gamma^i \frac{1}{{\not}D+m_s}    \right)
\end{equation}
with a trace over spin, color and space-time indices. To reduce the variance of $L(t)$ we rewrite it as~\cite{ETM:2008zte}
\begin{equation}
\label{eq:Lt2}
L(t) = \frac{1}{3}\mathrm{Tr} \left( \gamma^i \frac{m_s-m_l}{({\not}D+m_l)({\not}D+m_s)}   \right)
\end{equation}
for our calculation. This form makes explicit the cancellation of the quark-line disconnected contribution in the limit of equal quark masses, $m_l=m_s$. 

We use ensembles with approximate lattice spacing values of 0.15\,fm, 0.12\,fm and 0.09\,fm, Sets~1, 2 and~3 from Table~\ref{tab:ensembles}. The propagators are determined from stochastic random sources. On Set~3, and for some results on Set 2, we combine the truncated-solver method~\cite{Bali:2009hu, Alexandrou:2012zz} with deflation of low eigenmodes of the Dirac matrix~\cite{Wilcox:2007ei}. Early results were presented in Refs.~\cite{Yamamoto:2018cqm, FermilabLattice:2019dbx}. More recently the analysis has been blinded~\cite{FermilabLattice:2021hzx} to avoid bias in determination of the final value for the quark-line disconnected contribution to the total HVP. The blinding is done by multiplication of the correlators by a common unknown factor which is close to 1.

Since it is important not to unblind the full analysis prematurely, we use blinded data for our determination here of the disconnected contribution to $a_{\mu}^w$. To cancel the unknown blinding factor we must evaluate $a_{\mu}^w$ for this piece as a ratio to the full disconnected contribution to $a_{\mu}$ without a time-window. We do this on the lattices we have used for this calculation with smallest lattice spacing (Set~3 from Table~\ref{tab:ensembles} with $a \approx 0.09$\,fm), using preliminary results from 271 configurations~\cite{FermilabLattice:2021hzx}. This selection is sufficient for an estimate of the ratio in the continuum, when we allow a 15\% uncertainty for remaining discretisation effects. We then multiply this ratio by the BMW result for the disconnected contribution in the continuum limit and in infinite volume, after correcting our ratio for finite-volume and pion-mass effects using chiral perturbation theory. We take the BMW result to be $\big(-13.36(1.80)-2.1(3)\big)\times 10^{-10}$~\cite{Borsanyi:2020mff}), where the 2.1(3) is $-1/9$ of the complete finite-volume correction quoted, as appropriate for the $\pi\pi$ loop corrections~\cite{FermilabLattice:2019ugu}. This gives a total of $-15.46(1.82)\times 10^{-10}$, adding uncertainties in quadrature. 

In Sec.~\ref{sec:discwindow} we show that the window function with $t_1 \le 2$\,fm successfully removes the values of the disconnected correlation function with the largest statistical errors. This means that calculating $a_{\mu}^w$ from the disconnected correlation functions is both straightforward and precise. Here, however, because of the blinding, we must also calculate the full contribution in order to determine the ratio. This is harder to do. 

In order to reduce the uncertainty from large $t$ values in the full contribution to $a_{\mu}$ we adopt the strategy we have used for the connected case~\cite{Chakraborty:2016mwy, FermilabLattice:2019ugu}. We fit the correlators to a suitable functional form in $t$ and then use the fit function to calculate the contribution to $a_{\mu}$ from large $t$ values, rather than the correlator data. This allows the more precise small $t$ data to guide the values used at large $t$. The value of $t$ used at which we switch from using the correlation function data to using the fit results is called $t^*$. We can vary $t^*$ to tests for the stability of the results.

The fit function that we use for the disconnected correlation function is
\begin{eqnarray}
\label{eq:discfit}
G^{\mathrm{disc}}(t)&=&\sum_{i=1}^N \left[a_i^2e^{-E_{a_i}t} - b_i^2 e^{-E_{b_i}t}+\right.\\
&&\hspace{2.0em}\left.(-1)^t \left(c_i^2e^{-E_{c_i}t} -d_i^2e^{-E_{d_i}t}\right) \right] \, . \nonumber
\end{eqnarray}
This models the difference of isospin-1 and isospin-0 states that contribute to this correlation function~\cite{Chakraborty:2015ugp}. We take priors on the amplitudes, $a_i$, $b_i$, $c_i$ and $d_i$ in this fit to be $0\pm 0.1$. The prior for the ground-state ($\rho$) mass is taken as 0.3(1) and the difference between ground-state $\omega$ and $\rho$ masses is given the range 0.09(18). Excited-state energies are given priors of 0.5(4) above the mass below. 

Given the contribution from the time window and the total contribution, we can work out the ratio $a^w_{\mu}/a_{\mu}$ for the disconnected correlators. Multiplying this by the BMW result above gives the results for the disconnected correlator contributions in Table~\ref{tab:results}. The uncertainty in these is dominated by our determination of the full disconnected contribution, made necessary by the blinding. We stress that the determination of the disconnected contribution to $a_{\mu}^w$ using - results will be simpler and have smaller uncertainties (by at least a factor of 2) in the future. 

\end{appendix}

\bibliography{windows}

\end{document}